\documentclass[11pt, letterpaper]{article} 

\usepackage[T1]{fontenc}        
\usepackage[utf8]{inputenc}     
\usepackage{newtxtext, newtxmath} 
\usepackage{microtype}          
\usepackage[letterpaper, margin=1in, top=1in, bottom=1in]{geometry}
\usepackage[bottom]{footmisc}   
\usepackage{authblk}            

\usepackage{amsmath}
\usepackage{mathtools}
\usepackage{bbm}
\DeclareFontFamily{U}{bbm}{}
\DeclareFontShape{U}{bbm}{m}{n}
   {  <-> bbm10 }{}

\usepackage{graphicx}
\usepackage{subcaption}
\usepackage{tabularx}
\usepackage{booktabs}           

\usepackage{algorithm}
\usepackage{algpseudocode}

\usepackage[dvipsnames]{xcolor} 
\usepackage{soul}               

\usepackage[font=small,labelfont=bf,tableposition=top]{caption}

\usepackage[natbib=true, style=numeric-comp, sorting=none, sortcites=true, maxcitenames=2]{biblatex}
\usepackage{notoccite}
\usepackage{hyperref}
\hypersetup{
    colorlinks=true,
    linkcolor=BrickRed,
    citecolor=NavyBlue,
    urlcolor=NavyBlue,
    anchorcolor=NavyBlue,
    breaklinks=true
}

\title{\textbf{Accelerated Markov Chain Monte Carlo Simulation via \\ Neural Network--Driven Importance Sampling}}

\author[1]{Michael Kim}
\author[1]{Wei Cai\thanks{Email: \texttt{caiwei@stanford.edu}}}
\affil[1]{Department of Mechanical Engineering, Stanford University, CA 94305}

\date{\today}
\begin{document}
\maketitle

\begin{abstract}
%
%
%

Atomistic simulations provide valuable insights into the physical processes governing material behavior.
However, their applicability is fundamentally constrained by the limited time scales accessible to brute-force simulations. 
This bottleneck often stems from complex energy landscapes where the systems stay trapped in metastable states for long periods of time.
%
%
%
%
Yet, the long-term evolution is controlled by the transitions between the metastable states, which are rare events and difficult to observe.
%
%
%
%
%
%
We present an importance sampling method designed to accelerate the time scale of Markov chain Monte Carlo (MCMC) simulations. 
%
By employing a bias potential, our approach enhances the sampling of rare transition events while preserving the relative probabilities of distinct transition pathways.
%
The bias potential is represented by a neural network which enables the flexibility needed for high-dimensional systems.
%
%
We propose a rigorous formulation to obtain the original transition rates between metastable states using transition paths obtained from the biased simulation.
We further use a branching random walk (BRW) technique to enhance efficiency and to reduce variance.
The proposed methodology is validated on 2-dimensional and 14-dimensional systems, demonstrating its accuracy and scalability.
\end{abstract}

\section{Introduction}
\label{sec:introduction}
Atomistic simulations play a crucial role in understanding the fundamental processes governing the behavior of materials at the microscopic scale. 
However, they are fundamentally limited by the accessible time scale of brute-force methods.
Simulations frequently remain trapped within metastable states for extended periods, hindering the observation of transitions between these states.
Consequently, capturing these transitions that drive long-term evolution can be prohibitively expensive.
This intrinsic limitation is widely recognized as the time-scale or rare event problem~\cite{Voter2002,Voter200204,Voter2007}.

Various methodologies have been established to overcome the time-scale limitations of standard simulations. 
%
%
Methods such as hyperdynamics~\cite{Voter1997}, steered molecular dynamics~\cite{Izrailev1999}, and metadynamics~\cite{Parinello2002, Laio2008, Wang2018} rely on external forces or bias potentials to accelerate the escape from metastable basins.
Alternatively, trajectory-based approaches, including the weighted ensemble method~\cite{huber1996weighted, zuckerman2017weighted} and forward flux sampling~\cite{allen2009forward}, enhance sampling efficiency by selectively branching trajectories along a reaction coordinate. 
In parallel, transition path theory~\cite{Dellago2002, vanden-eijnden2006, E2006, metzner2009transition} offers a mathematical basis for analyzing the statistics of reactive pathways. 
Importance sampling methods~\cite{Kahn1953, Glynn1989, Cai2002, Ecuyer2009, Dupuis2004, de2005adaptive} represent another established direction, designed to capture dominant transition pathways and estimate dynamical properties of rare events.
In this work, we focus on applying importance sampling to extend the time scale of Markov chain Monte Carlo simulations.


The importance sampling method pursues two primary objectives.
First, it aims to enhance the sampling of successful transition paths between metastable states without artificially favoring one reaction channel over another.
Specifically, the relative probabilities of competing pathways are preserved to ensure the transition mechanism remains unbiased.
Second, the applied enhancement needs to be quantifiable, allowing the true transition rates of the original system to be recovered.
In practice, importance sampling modifies the system kinetics by updating transition probabilities with an importance function~\cite{Cai2002, de2005adaptive}.
The requirement that these updated probabilities remain normalized serves as a constraint that uniquely determines the optimal importance function.
Consequently, importance sampling reformulates the rare event problem into an optimization problem.

Several significant challenges still exist for applying the importance sampling method to address the time scale problem in atomistic simulations.
%
An increase in the size or dimensionality of the system can drastically increase the computational burden of finding the optimal importance function. 
In addition, numerical stability becomes critical at low temperatures, where the magnitude of the optimal importance function near the energy minima of the starting metastable state can become exceedingly small, leading to underflow errors. 
Most importantly, the accuracy of the estimated transition rates relies heavily on the quality of the importance function.
Even a small deviation from the optimal solution causes the variance of the estimator to increase drastically, resulting in large errors in the estimated transition rates.


In this work, we propose a robust method to overcome these limitations in Markov chain Monte Carlo (MCMC) simulations of spatially discrete systems. 
First, to address the challenge of searching for the optimal importance function in high dimensions, we employ a neural network combined with adaptive sampling. 
Second, to resolve numerical instability at low temperatures, we perform the optimization in the logarithmic space by introducing an optimal bias potential, which effectively prevents underflow issues near energy minima. 
Finally, we present a general formulation for estimating transition rates, complemented by a branching random walk (BRW) technique to significantly reduce the variance of the estimator.

The remainder of this paper is organized as follows. 
Section~\ref{sec:prob_def} formulates the problem and defines the objectives of the importance sampling framework. 
Section~\ref{sec:methods} details the proposed method, covering the generalized formulation, the transition rate estimation technique, and the neural network training protocol. 
Section~\ref{sec:results} demonstrates the efficacy of the method through numerical experiments on both 2-dimensional and 14-dimensional systems. 
Finally, Section~\ref{sec:conclusion} summarizes our findings and discusses potential directions for future research.

\section{Problem Statement}
\label{sec:prob_def}
\begin{figure*}[htbp!]
\centering
\includegraphics[width=.40\textwidth]{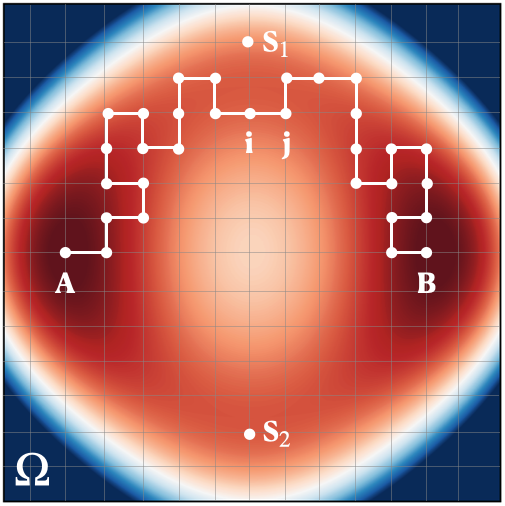}%
        \label{subfig:pel_discrete_omega}%
    \caption{Domain $\Omega$ is a set of allowed microscopic states (e.g. labeled as $i$ and $j$) correspond to points on a regular grid in a two-dimensional region. The colored background corresponds to the energy landscape $E(i)$. 
    Microscopic states A and B corresponding to the local minima of the energy landscape. Microscopic states S$_1$ and S$_2$ are the saddle points of the energy landscape. The white path through the microscopic states indicates a transition path from A to B.}
    \label{fig:pel_discrete}
\end{figure*}
\noindent Consider a domain $\Omega$, which is a set of allowed microscopic states corresponding to points on a regular grid in a two-dimensional region, as shown in Figure~\ref{fig:pel_discrete}.
On this domain, an energy landscape $E(i)$ is defined, which specifies the energy of any microscopic state $i$.
The kinetics of this system is defined through function $r(i \to j)$, which is the jump rate between any two microscopic states $i$ and $j$ on the grid,
\begin{equation}
    r(i \rightarrow j) = \nu_0 \, \exp \left[ - \frac{E(j) - E(i)}{2\,k_{\textrm{B}}T}\right] \, \mathbbm{1}_{\textrm{nn}}(i,j) \, ,
    \label{eq:transition_rate_single}
\end{equation}
%
%
\noindent where $\nu_0$ is a constant (frequency prefactor), and $\mathbbm{1}_{\textrm{nn}}(i,j)$ equals 1 if $i$ and $j$ are nearest neighbor grid points in $\Omega$ and 0 otherwise. 
It can be easily shown that the jump rate defined in Eq.~(\ref{eq:transition_rate_single}) satisfies the detailed balance condition,
%
%
\begin{equation}
r(i\to j) \, \exp\left[-\frac{E(i)}{k_{\rm B}T} \right] = 
r(j\to i) \, \exp\left[-\frac{E(j)}{k_{\rm B}T} \right] \, , \quad 
\forall \, i, j \in \Omega \, .
\end{equation}
Assume that the energy landscape contains two deep basins, corresponding to two metastable states.
Let A and B be the (local) energy minima at the bottom of these two basins. 
Assume that there are two saddle points, S$_1$ and S$_2$, located on two distinct transition pathways between A and B.
At low temperatures, if the system starts in the neighborhood of A, it is likely to spend a long time in the basin containing A, and the transition to the basin containing B is a rare event.
When such a rare transition does occur, the transition path is likely to pass through the neighborhood of either S$_1$ or S$_2$.

Importance sampling aims to accelerate the time scale of MCMC simulations, while preserving the relative probabilities of distinct pathways. 
The objectives are twofold: (1) obtain accurate transition rates between metastable states and (2) reveal the underlying transition mechanisms by sampling transition paths with their correct probabilities.
Although the example used above is a simple 2-dimensional system, the algorithm is designed to be robust enough to handle high-dimensional systems, where rare events are computationally prohibitive.
%
%


\section{Methods}
\label{sec:methods}
\subsection{Importance Sampling of Rare Transition Paths}
\label{subsec:system1}
Let us start by taking a closer look at the jumps between the microscopic states. 
If the system is in microscopic state $i$, the probability of the jumping to microscopic state $j$ upon leaving $i$ is,
\begin{equation}
    K(i \rightarrow j) = \frac{r(i \rightarrow j)}{r_{\rm tot}(i)}, \hspace{5mm} r_{\rm tot}(i) = \sum_{k \in \mathcal{N}(i)} \, r(i \rightarrow k),
    \label{eq:prob_rate}
\end{equation}
where $r_{\rm tot}(i)$ is the total jump rate out of microscopic state $i$ and $\mathcal{N}(i)$ is the set of all microscopic states accessible from (i.e. adjacent to) microscopic state $i$.

We now define a path as a Monte Carlo trajectory that starts at A and ends at either A or B. 
Furthermore, we shall call paths that end at A the ``failure'' paths, and paths that end at B the ``success'' paths.  
For example, the probability of sampling a given ``success'' path
$\Gamma_{\rm AB}=\{\textrm{A}, i_1, i_2, \cdots, i_n, \textrm{B}\}$ is,
\begin{equation}
  P (\Gamma_{\rm AB}) = K(\textrm{A} \to i_1) \cdot K(i_1 \to i_2) \cdots K(i_n \rightarrow \textrm{B}).
\end{equation}
The rare event problem stems from the fact that, under the dynamics defined by $K(i \to j)$, the probability of sampling a ``success'' path is extremely small. 
Consequently, direct Monte Carlo simulations are computationally inefficient, as the vast majority of resources are expended sampling ``failure'' paths.

With importance sampling, we aim to construct a new simulation (kinetics), where the jump probability $K'$ between two microscopic states is modified from the original probability $K$ as follows,
\begin{equation}
    K'(i \to j) = K(i \to j) \, \frac{I(j)}{I(i)} \, \Big[1 - \mathbbm{1}_{\textrm{A}}(j) \Big],
    \label{eq:prob_update_ideal}
\end{equation}
where $I(i)$ is called the \emph{importance function}, and $\mathbbm{1}_{\textrm{A}}(j) = 1$ if $j$ = A and $\mathbbm{1}_{\textrm{A}}(j) = 0$ otherwise.
However, Eq.~(\ref{eq:prob_update_ideal}) is not always possible, because $K'(i\to j)$, being a probability matrix, has to be normalized,
\begin{equation}
  \sum_{j \in \mathcal{N}(i)} K'(i\to j) = 1.
  \label{eq:normalize_k_prime}
\end{equation}
We define the importance function that guarantees the normalization condition, Eq.~(\ref{eq:normalize_k_prime}), as the optimal importance function, $I^{\rm opt}(i)$.
In general, when $I(i)$ does not equal to $I^{\rm opt}(i)$ (because the latter is unknown), we must introduce a normalization factor, $n'(i)$, to make sure that $K'(i\to j)$ is properly normalized.
\begin{equation}
    K'(i \to j) = \frac{1}{n'(i)} \, K(i \to j) \, \frac{I(j)}{I(i)} \, \Big[1 - \mathbbm{1}_{\textrm{A}}(j) \Big], \quad 
    n'(i) = \sum_{j \in \mathcal{N}(i)} \, K(i \to j) \, \frac{I(j)}{I(i)} \, \Big[1 - \mathbbm{1}_{\textrm{A}}(j) \Big].
    \label{eq:new_trans_prob}
\end{equation}
The new probability of sampling the same path, $\Gamma_{\rm AB}$, in the modified Monte Carlo simulation is,
\begin{equation}
  P' (\Gamma_{\rm AB}) = K'(\textrm{A} \to i_1) \cdot K'(i_1 \to i_2) \cdots K'(i_n \rightarrow \textrm{B})
  = \left[ \frac{1}{n'(\textrm{A})} \, \prod_{k=1}^{n} \frac{1}{n'(i_k)} \right] \, P (\Gamma_{\rm AB}) \, \frac{I(\textrm{B})}{I(\textrm{A})}.
\end{equation}
In the event that all the normalization factors equal to $1$ (when $I(i)=I^{\rm opt}(i)$), we would have achieved a remarkable outcome,
\begin{equation}
  P' (\Gamma_{\rm AB}) =  P (\Gamma_{\rm AB}) \, \frac{I^{\rm opt}(\rm B)} {I^{\rm opt}(\rm A)}.
\end{equation}
This expression implies that the two requirements identified in Section~\ref{sec:prob_def} are exactly satisfied: (1) the sampling of all successful paths from A to B are enhanced by the same amount (i.e. the enhancement factor is path independent), and (2) the enhancement factor is known, so that we can obtain the transition rate of the original system from that of the modified system.

In practice, the optimal importance function $I^{\rm opt}(i)$ is not known, and finding it exactly is at least as difficult as solving the original rare event problem.
The well-established approach of importance sampling (e.g. widely successful in quantum chemistry~\cite{kalos2009monte}) is to develop ``good enough'' approximations to $I^{\rm opt}(i)$.
What has been achieved in the above is the reformulation of the rare event (or time scale) problem into an optimization problem, in search for the optimal importance function.
The optimality condition for the importance function is given by Eq.~(\ref{eq:normalize_k_prime}), or equivalently,
\begin{equation}
 \sum_{j \in \mathcal{N}(i) \setminus \{ {\rm A} \} } K(i\to j) \, I(j) = I(i).
 \label{eq:eigen_K_I}
\end{equation}
In other words, the optimal importance function is the (right) eigenvector of the transition probability matrix $K(i\to j) \Big[1 - \mathbbm{1}_{\textrm{A}}(j) \Big] $ with eigenvalue $1$ (which corresponds to the largest eigenvalue).
We set $I(\textrm{B}) = 1$ without loss of generality.
The optimal importance function corresponds to the discrete committor function (see Section~\ref{sec:rate_estimate}).

Recalling the jump rate expression between microscopic states, Eq.~(\ref{eq:transition_rate_single}), it can be shown that modifying the jump probability matrix using the importance function $I(i)$, as in Eq.~(\ref{eq:prob_update_ideal}), is equivalent to introducing a bias potential, 
\begin{equation}
  E_{\rm b}(i) = -2 \, k_{\rm B} T \log I(i).
  \label{eq:bias_pot}
\end{equation}
which is added to the original potential energy $E(i)$ to modify the Monte Carlo trajectory.
Searching for the optimal importance function is thus equivalent to searching for the optimal bias potential.
\subsection{Generalized Formulation of Importance Sampling of Rare Events}
\label{subsec:system2}

For simplicity, the system described in Section 2.1 is defined on a discretized spatial grid, with grid spacing $\Delta x$ in every dimension.  While convenient, this discretization becomes an artifact when modeling atoms whose positions can be any real number.
This artifact will vanish in the continuum limit where the grid spacing, $\Delta x$, approaches zero. 
However, the above formulation encounters a significant problem when the grid spacing is systematically refined: the ``failure'' or ``success'' designation of a trajectory is defined through discrete microscopic states (A or B), which serve as boundary conditions for the optimal importance function.
As the grid spacing is systematically refined, the solution to the optimal importance function exhibits an increasingly sharp gradient around the microscopic points A and B.
This behavior is analogous to applying boundary conditions to a single point in a continuum satisfying a partial differential equation (PDE), which is generally ill-advised. 
A more robust formulation, as presented below, involves defining ``failure'' or ``success'' of a trajectory using a finite region rather than a single point.
The size of such regions remains constant as the discretization grid size is refined.

Instead of selecting two microscopic states, A and B, and designate them as ``failure'' and ``success'', we now augment the original system (in domain $\Omega$) with two additional states, F and S.
In other words, we enlarge the domain to $\tilde{\Omega} = \Omega \cup \{ {\rm F}, {\rm S} \}$.
We now modify the kinetics of the original system so that multiple microscopic states (around A or B) have a nonzero jump rate into F or S.
If a trajectory enters state F, it is called a ``failure'', and if a trajectory enters state S, it is called a ``success''.

We introduce functions $f(i)$ and $s(i)$ to control whether microstate $i$ is allowed to jump into F or S (only if $f(i)$ or $s(i)$ is non-vanishing).
In practice, we shall make $f(i)$ non-zero only in a small area around the microscopic state A, and $s(i)$ non-zero only in a small area around the microscopic state B.
For example, the functions $f(i)$ and $s(i)$ can be selected to have a form of a truncated Gaussian as follows.
\begin{align}
    f(i) &= \epsilon \, (\Delta x)^d \, \exp \left( -\frac{\|\textbf{x}(i) - \textbf{x}(\textrm{A})\|^2}{2\sigma^2} \right) \cdot \mathbbm{1} \left( \|\textbf{x}(i) - \textbf{x}(\textrm{A})\| < r_{\rm c}\right) \, ,
    \label{eq:control_functions_f}\\
    s(i) &= \epsilon \, (\Delta x)^d \, \exp \left( -\frac{\|\textbf{x}(i) - \textbf{x}(\textrm{B})\|^2}{2\sigma^2} \right) \cdot \mathbbm{1} \left(\|\textbf{x}(i) - \textbf{x}(\textrm{B})\| < r_{\rm c}\right) \, ,
    \label{eq:control_functions_s}
\end{align}
%
%
where $\textbf{x}(i)$ is the position vector of microscopic state $i$ in a $d$-dimensional system, $\|\cdot\|$ represents the Euclidean norm, and $\mathbbm{1}(\cdot)$ is an indicator function that equals 1 if the condition inside the parentheses is satisfied and equal 0 otherwise.
%
%
The prefactor $(\Delta x)^d$ is introduced to ensure that the flux between state $i$ and F remains nearly constant regardless of the choice of $\Delta x$.
%
%

To fully specify the kinetics of the extended system (in domain $\tilde{\Omega}$), we also define the energy of the two auxiliary microscopic states, $E(\textrm{F})$ and $E(\textrm{S})$.
Then the jump rate between any two microscopic states $i$ and $j$ in $\tilde{\Omega}$ ($i \neq j$) can be written as,
\begin{equation}
    \tilde{r}(i \rightarrow j) = \nu \, \exp \left[ - \frac{E(j) - E(i)}{2\,k_{\textrm{B}}T}\right]
    \cdot
    \Big[
    \mathbbm{1}_{\textrm{nn}}(i,j) + 
    \mathbbm{1}_{\textrm{F}}(j) \, f(i) +
    \mathbbm{1}_{\textrm{F}}(i) \, f(j) +
    \mathbbm{1}_{\textrm{S}}(j) \, s(i) +
    \mathbbm{1}_{\textrm{S}}(i) \, s(j) 
    \Big],
    \label{eq:transition_rate_extended}
\end{equation}
where $\mathbbm{1}_{\textrm{F}}(j) = 1$ if $j = \textrm{F}$ and $\mathbbm{1}_{\textrm{F}}(j) = 0$ otherwise, and $\mathbbm{1}_{\textrm{S}}(j) = 1$ if $j = \textrm{S}$ and $\mathbbm{1}_{\textrm{S}}(j) = 0$ otherwise.
$\tilde{r}(i \rightarrow j) = 0$ if $i = j$.
The jump rate defined in this way satisfies the detailed balance condition, similar to $r(i\to j)$,
\begin{equation}
\tilde{r}(i\to j) \exp\left[-\frac{E(i)}{k_{\rm B}T} \right] = 
\tilde{r}(j\to i) \exp\left[-\frac{E(j)}{k_{\rm B}T} \right] \, , \ \ 
{\rm for \ any }  \ \ i, j \in \tilde{\Omega} \, .
\end{equation}
The goal of this design is to minimize the perturbation of the kinetics of the original system (defined by $r(i\to j)$ on $\Omega$).
All we have added are two extra microscopic states (F and S); the effect should be minimal when the total number of states in $\Omega$ becomes very large, e.g. in the limit of grid spacing going to zero.

Given the jump rates, we can define the probability of jumping to microscopic state $j$ upon leaving $i$, similar to before, as follows.
\begin{equation}
    \tilde{K}(i \rightarrow j) = \frac{\tilde{r}(i \rightarrow j)}{\tilde r_{\rm tot}(i)}\, , \hspace{5mm} \tilde  r_{\rm tot}(i) = \sum_{j \in \mathcal{A}(i)} \tilde{r}(i \rightarrow j),
\end{equation}
where $\mathcal{A}(i) = \mathcal{N}(i) \cup \{ {\rm F}, {\rm S}\} $ is the set of microscopic states that the system is allowed to jump from $i$.

The addition of the two new microscopic states, F and S, is expected to bring minimal changes to the kinetics of the original system.
Therefore, we expect the kinetics of the extended system to still exhibit the characteristics of rare events.  If the system is initially in the neighborhood of microscopic state A, it tends to spend a long time in this region before a rare transition occurs that brings it to the neighborhood of microscopic state B.
Because $f(i) > 0$ only in a small region around A and $s(i) > 0$ only in a small region around B, the system repeatedly visits F for many times before it visits S for the first time.
We define a trajectory that starts and ends at F as a ``failure'' path, and a trajectory that starts at F and ends at S a ``success'' path.
Then the rare event problem is equivalent to the extremely low probability of sampling ``success'' paths using the transition probability $\tilde{K}(i\to j)$.

To use importance sampling to overcome the rare event problem, we introduce importance function $I(i)$ defined on every $i \in \Omega$ to modify the transition probability as follows.
\begin{equation}
    \tilde K'(i \to j) = \frac{1}{\tilde{n}'(i)} \, \tilde{K}(i \to j) \, \frac{I(j)}{I(i)} \, \Big[1 - \mathbbm{1}_{\textrm{F}}(j) \Big], \quad 
    \tilde{n}'(i) = \sum_{j \in \mathcal{A}(i)} \, \tilde{K}(i \to j) \, \frac{I(j)}{I(i)} \, \Big[1 - \mathbbm{1}_{\textrm{F}}(j) \Big],
    \label{eq:new_trans_prob_ext}
\end{equation}
where $\tilde{n}'(i)$ is the normalization factor. The optimality condition is when the normalization factors $\tilde{n}'(i) = 1$ for all $i \in \Omega$, which can be equivalently written as follows.
\begin{equation}
 \sum_{j \in \mathcal{A}(i) \setminus \{ {\rm F}\}} \tilde{K}(i\to j) \, I(j) = I(i).
 \label{eq:eigen_K_I_gen}
\end{equation}
Note that $\mathcal{A}(i) \setminus \{ {\rm F}\} = \mathcal{N}(i) \cup \{\textrm{S}\}$, i.e. the sum over $j$ does not include state F. 
Also note that in this formulation, importance sampling is not applied to jumps out of state F, i.e.,
\begin{equation}
\tilde K'({\rm F} \to j) = \tilde K({\rm F} \to j) \, ,
\end{equation}
so that it is not necessary to define $I(\rm {F})$.
In the following, we shall set $I(\textrm{S}) = 1$ without loss of generality.

\subsection{Transition Rate Estimation}
\label{sec:rate_estimate}

\begin{figure*}[htbp!]
\centering
    \includegraphics[width=.84\textwidth]{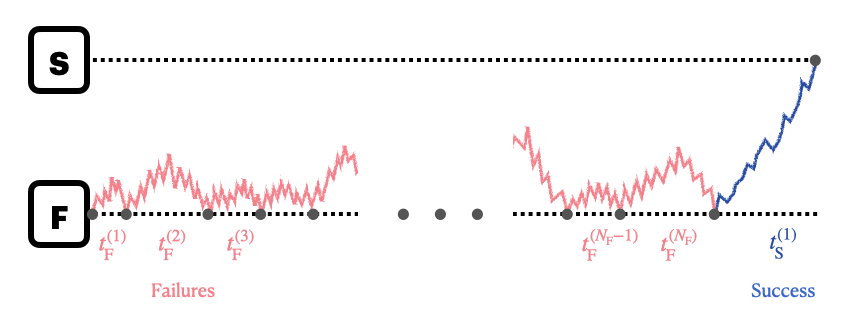}%
    
    \caption{Transition from state F to S. Failure (i.e., coming back to F after starting from F) occurs $N_{\textrm{F}}$ times before the success (i.e., arriving at S before hitting F after starting from F) takes place. The first passage time can be considered as a sum of the duration of failure and success paths. In the case of rare events, the number of failures before success is very large.}
    \label{fig:tran_rate_est}
\end{figure*}
Here we describe the method for computing the transition rates between metastable states using importance sampling, based on the generalized formulation introduced in Section~\ref{subsec:system2}. 
These results can be readily adapted to the simplified formulation of Section~\ref{subsec:system1} by substituting F and S with A and B, respectively.

With the definition of the auxiliary microscopic states F and S, the transition rate between the two metastable states is very close to the transition rate between the microscopic states F and S,
$r_{\rm FS}$, which equals to the inverse of the mean first passage time $T_{\rm FS}$ from F to S.
This implies,
\begin{equation}
r_{\rm FS} = \frac{1}{\langle T_{\rm FS}\rangle}
\end{equation}
where $\langle \cdot \rangle$ indicates average over all statistical realizations.
Consider a trajectory that starts from F and eventually ends at S as shown in Figure~\ref{fig:tran_rate_est}.  In a typical situation, the system visits F many times before visiting S for the first time. 
Therefore, the mean first passage time reads,
\begin{equation}
\langle T_{\rm FS} \rangle = \langle N_{\rm F} \rangle \langle t_{\rm FF} \rangle + \langle t_{\rm FS} \rangle
\end{equation}
where $\langle N_{\rm F} \rangle$ is the average number of times the system returns to F before visiting S, $\langle t_{\rm FF} \rangle$ is the average time of a ``failure'' path (starting from F and ending at F), and $\langle t_{\rm FS} \rangle$ is the average time of a ``success'' path (starting from F and ending at S).
In the evaluation of $\langle t_{\rm FF} \rangle$, we adopt the convention of excluding the time accrued 
while the system stays in state F (before making a transition to a microstate $i \in \Omega$).
%
%
The purpose of this convention is to minimize the perturbation to the original system, so that the transition rate $r_{\rm FS}$ follows the transition rate between metastable states of the original system as much as possible.
%
In the rare event limit where the transition between metastable states is very rare, $\langle N_{\rm F}\rangle \gg 1$,
\begin{equation}
\langle T_{\rm FS} \rangle \approx \langle N_{\rm F} \rangle \langle t_{\rm FF} \rangle
\approx \frac{ \langle t_{\rm FF} \rangle }{p_{\rm S}({\rm F}) }
\label{eq:T_FS_pS}
\end{equation}
where $p_{\rm S}({\rm F})$ is the probability of the system starting from F to visit S before returning to F.
In general, define $p_{\rm S}(i)$ as the probability of the system starting from microscopic state $i$ to visit S before visiting F.
$p_{\rm S}(i)$ is often referred to as the discrete \textit{committor function} and equals to the optimal importance function $I^{\rm opt}(i)$.
Therefore, the transition rate can be written as,
\begin{equation}
r_{\rm FS} 
\approx \frac{p_{\rm S}({\rm F}) } { \langle t_{\rm FF} \rangle }
= \frac{I^{\rm opt}({\rm F}) } { \langle t_{\rm FF} \rangle }
\label{eq:s_FS_pS}
\end{equation}
where $I^{\rm opt}({\rm F}) = \sum_{j \in \Omega} \tilde{K}(\textrm{F} \to j) \, I^{\rm opt}(j)$.
It is straightforward to compute $\langle t_{\rm FF} \rangle$ using the Markov chain Monte Carlo algorithm~\cite{voter2007introduction, jansen2012introduction, battaile2008kinetic} with $\tilde{r}(i,j)$ without importance sampling.
These simulations can easily sample a large number of ``failure'' paths to determine their average duration, $\langle t_{\rm FF} \rangle$.
Given Eq.~(\ref{eq:s_FS_pS}), the transition rate $r_{\rm FS}$ can be readily obtained if the optimal importance function $I^{\rm opt}(i)$ is known.
In practice, $I^{\rm opt}(i)$ is often unknown, and we have to rely on an approximation, $I(i)$.
To estimate the transition rate $r_{\rm FS}$, our primary task is to determine $p_{\rm S}(\textrm{F})$ using this approximate importance function $I(i)$.

From the problem statement in Section~\ref{sec:prob_def}, it is clear that the transition rate $r_{\rm FS}$ is proportional to the frequency prefactor $\nu_0$ that appears in the definition of jump rate between any two microscopic states, Eq.~(\ref{eq:transition_rate_single}).
At the same time, the grid spacing $\Delta x$ is often considered a numerical artifact, and it is of interest to consider the continuum limit of $\Delta x \to 0$.
Therefore, it is necessary to consider how the frequency prefactor $\nu_0$ should change when the grid spacing $\Delta x$ varies.
In this work, we let the frequency prefactor $\nu_0$ depend on temperature and grid spacing as follows.
\begin{equation}
\nu_0 = \frac{\mu \, k_{\rm B}T}{(\Delta x)^d},
\label{eq:nu_0_mu_kBT_Dx2}
\end{equation}
where $\mu$ is the mobility constant (we choose $\mu = 1.0\,\mathrm{m}^2\,\mathrm{s}^{-1}\,\mathrm{eV}^{-1}$), $k_{\rm B}$ is the Boltzmann constant, $T$ is the temperature, and $d$ is the dimension.
This allows the transition rate $r_{\rm FS}$ to remain unchanged in the limit of $\Delta x \to 0$ (see Figure~\ref{subfig:rate_nn_conv_annh}).
\subsection{Success Probability Calculation}
\label{sec:success_prob}

We now discuss how to compute the success probability $p_{\rm S}(\textrm{F})$, which is the probability of a trajectory starting at F to visit S before visiting F again, given an importance function $I(i)$ that is not necessarily the optimal importance function.
For this purpose, let us define a path $\Gamma$ as a trajectory of the system that starts at F and ends at either F or S, without visiting either F or S in between.
Let $P(\Gamma)$ be the probability of sampling path $\Gamma$ following transition probability $\tilde{K}(i,j)$.
Furthermore, let us define $\{\Gamma_{\rm FF}\}$ as the set of paths that starts at F and ends at F; these are the ``failure'' paths.
Let us define $\{\Gamma_{\rm FS}\}$ as the set of paths that starts at F and ends at S; these are the ``success'' paths.
With these definitions, we can write the success probability $p_{\rm S}({\rm F})$ as,
\begin{equation}
  p_{\rm S}(\textrm{F}) = \sum_{\Gamma \in \{ \Gamma_{\rm FS}\}} P(\Gamma)
\end{equation}
Let us define a ``reward'' function $R$ for each path $\Gamma$ such that
\begin{equation}
 R(\Gamma) = \left\{ 
     \begin{array}{ll}
       1 \, , & {\rm if } \ \Gamma \in \{ \Gamma_{\rm FS} \} \\
       0 \, , & {\rm otherwise} 
     \end{array}
 \right.
\end{equation}
Then we can rewrite the success probability as,
\begin{equation}
  p_{\rm S}({\rm F}) = \sum_{\Gamma} R(\Gamma) \, P(\Gamma)
  \label{eq:PS_F_sum_over_all_paths}
\end{equation}
where the sum is over all paths (regardless of success or failure).
This means that $p_{\rm S}({\rm F})$ can also be expressed as a statistical average over all paths sampled by the Monte Carlo simulation,
\begin{equation}
p_{\rm S}({\rm F}) = \langle R(\Gamma) \rangle_{\Gamma}
\label{eq:PS_F_as_path_average}
\end{equation}
where $\langle \cdot \rangle_{\Gamma}$ means averaging over all paths sampled by the Monte Carlo simulation using transition probability $\tilde{K}(i,j)$.
Even though Eq.~(\ref{eq:PS_F_as_path_average}) gives a very simple recipe to compute the success probability, for systems that exhibit rare event behaviors, it can be extremely inefficient.
This is because the vast majority of the paths sampled are ``failure'' paths for which the ``reward'' is zero.
The importance sampling approach is designed to address this inefficiency problem.

We now use the importance function $I(i)$ to modify the transition probability to $\tilde{K}'(i\to j)$, as defined in Eq.~(\ref{eq:new_trans_prob_ext}).
Consider a specific ``success'' path $\Gamma_{\rm FS} = \{{\rm F}, i_1, i_2, \cdots, i_n, {\rm S}\}$.
The new probability $P'(\Gamma_{\rm FS})$ of sampling this path using the importance function is related to the original probability $P(\Gamma_{\rm FS})$ as follows.
\begin{equation}
  P' (\Gamma_{\rm FS}) =
 \left[ \prod_{k=1}^{n} \frac{1}{\tilde{n}'(i_k)} \right] \, P (\Gamma_{\rm FS}) \, \frac{I(\textrm{S})}{I(i_1)}.
\end{equation}
where $\tilde{n}'(i)$ is normalization factor defined in Eq.~(\ref{eq:new_trans_prob_ext}).
Define the weight of path $\Gamma$,
\begin{equation}
W(\Gamma_{\textrm{FS}}) \coloneqq \prod_{k=1}^{n} {\tilde{n}'(i_k)} 
\end{equation}
as the product of the normalization factors of all states visited along the path excluding the starting and final states (F and S).
Then, the original probability can be written as,
\begin{equation}
  P (\Gamma_{\rm FS}) = W(\Gamma_{\rm FS})
  \, \frac{I(i_1)}{I(\textrm{S})} \, P' (\Gamma_{\rm FS}),
\end{equation}
which can be substituted into Eq.~(\ref{eq:PS_F_sum_over_all_paths}) to obtain,
\begin{equation}
p_{\rm S}({\rm F}) = \frac{1}{I(\textrm{S})}\, \sum_{\Gamma} R(\Gamma) \, W(\Gamma) \, I(i_1) \, P' (\Gamma) 
= \frac{1}{I(\textrm{S})} \,
\Big\langle R(\Gamma) \, W(\Gamma) \, I(i_1) \Big\rangle_{\Gamma}' 
\label{eq:PS_F_by_imp_sample_R_W}
\end{equation}
where $\langle \cdot \rangle'_{\Gamma}$ means averaging over all paths sampled by the Monte Carlo simulation using transition probability $\tilde{K}'$, i.e. using importance sampling.
The advantage of importance sampling is that, if $I(i)$ is close to $I^{\rm opt}(i)$, then it is very likely to sample ``success'' paths for which the ``reward'' is non-zero.
If $I(i)$ is equal to $I^{\textrm{opt}}(i)$, the success probability expression reduces to,
\begin{equation}
    p_{\textrm{S}}(\textrm{F}) = \sum_{j \in \Omega} \tilde{K}(\textrm{F} \rightarrow j) \, I^{\rm opt}(j).
    \label{eq:PS_F_no_imp_sample_R_W}
\end{equation}
This means that the success probability can be computed without sampling any transition paths if $I^{\rm opt}(i)$ is known.
However, since in practice the importance function $I(i)$ accessible to us is only an approximation of $I^{\rm opt}(i)$, the proper way to compute the success probability is through Eq.~(\ref{eq:PS_F_by_imp_sample_R_W}), which requires importance sampling of transition paths.
%
%
%
Recall that the first step of every path, i.e. the jump out of state F, is not subjected to importance sampling, i.e. $\tilde K'({\rm F} \to j) = \tilde K({\rm F} \to j)$.
The details of how to implement this jump in high-dimensional systems is described in Section~\ref{subsec:determine_initial_state}.
\subsection{Efficient Sampling via Branching Random Walk}
\label{sec:brw_method}

While Eq.~(\ref{eq:PS_F_by_imp_sample_R_W}) provides an unbiased estimate of the success probability, $p_{\textrm S}({\textrm F})$, it is still necessary to sample a sufficient number of paths to achieve statistical convergence.
The number of paths required to reduce the statistical error to an acceptable level is highly dependent on the quality of the importance function, $I(i)$, used for sampling.
Unfortunately, in practice, $I(i)$ may be so different from $I^{\rm opt}(i)$ that an excessive number of paths is needed if Eq.~(\ref{eq:PS_F_by_imp_sample_R_W}) were used directly to estimate $p_{\rm S}({\rm F})$, undermining the efficiency of the importance sampling method.
This problem can be addressed by replacing the importance sampled Monte Carlo simulation with a branching random walk (BRW) process.

The BRW process starts from a single random walker located at state F, where its weight is set to unity. 
At each step of the BRW process, the random walker updates its weight, by multiplying it with the normalization factor of its current state. 
Then, the random walker can either annihilate or split into multiple random walkers, depending on its weight.
For example, if the weight, $W$, is less than a threshold $W_{\rm min}$ (e.g., 0.5), an attempt is made to annihilate the walker with probability $1 - W$. 
If the walker is annihilated, the path is terminated and marked as ``failure'', i.e., the ``reward'' is zero. 
If the walker survived the annihilation attempt, its weight $W$ is restored to 1.
On the other hand, if the weight $W$ is greater than a threshold $W_{\rm max}$ (e.g., 2), the walker is split into $\lfloor W \rfloor + 1$ walkers with probability $W - \lfloor W \rfloor$, or $\lfloor W \rfloor$ walkers otherwise.
The weight of the walkers are reset to 1.
All the walkers remaining after the annihilation and splitting process travel in the domain $\Omega$ following the same transition probabilities $\tilde{K}'$, and may annihilate or further split in the next steps, depending on their weights.
If any walker visits state S, it is terminated and its path is marked as ``success''.
The process continues until all the walkers are terminated either by annihilation or having reached state S.

The BRW process can be rigorously formulated using a stochastic rounding function, a mathematical technique widely applied in Monte Carlo rendering and particle transport simulations~\cite{vorba2016adjoint, arvo1990particle, rath2022ears}.
The stochastic rounding function, $\mathcal{R}(W)$, can be written as follows.
\begin{equation}
\mathcal{R}(W) = \begin{cases}
\lfloor W \rfloor + 1 & \text{with probability } W - \lfloor W \rfloor \cr
\lfloor W \rfloor & \text{otherwise.}
\end{cases}
\end{equation}
At every step of the BRW process, if a random walker's weight falls outside the threshold bounds, $[W_{\rm min}, \, W_{\rm max}]$, the random walker transforms into $\mathcal{R}(W)$ new walkers.
For example, $\mathcal{R}(W) = 0$ implies that the original walker is terminated, while $\mathcal{R}(W) \geq 1$ indicates that the process proceeds with $\mathcal{R}(W)$ walkers.
This process is repeated until all walkers have either been terminated by annihilation or having reached state S.

%
Once the BRW process is complete, any walker marked as a ``success'' receives a base reward of unity, while those marked as a ``failure'' receive zero.
To compute the total score, the reward of each successful walker is weighted by the product of its current path weight, $W(\Gamma)$, and $I(i_1)$, the value of the importance function at the first intermediate state.
This weighted summation provides a statistical estimate of the quantity $\langle R(\Gamma) \, W(\Gamma) \, I(i_1) \rangle_{\Gamma}'$, ensuring the result is unbiased.
%
Meanwhile, the branching mechanism ensures that the path weights, $W(\Gamma)$, remain controlled within the prescribed bounds $[W_{\rm min}, \, W_{\rm max}]$.
Consequently, this process enables an efficient and accurate estimation of the success probability, $p_{\textrm{S}}(\textrm{F})$, and by extension, the transition rate, $r_{\textrm{FS}}$.
\subsection{Neural Network Representation of Bias Potential}
\label{sec:nn_bias_pot_train}

When the grid spacing of the system is systematically refined, the total number of microscopic states increases. 
This increase is drastic in high-dimensional systems.
%
Therefore, it becomes computationally intractable to optimize the importance function, $I(i)$, or the bias potential, $E_{\textrm{b}}(i)$, for every microscopic state in high-dimensional systems.

To overcome this limitation, we integrate machine learning techniques~\cite{li2022semigroup, yuan2024optimal, mitchell2024} into the importance sampling framework. 
The importance function $I(i)$ can be represented by a neural network as $I(i;\,\theta)$, where $\theta$ are trainable parameters.
The neural network importance function can be optimized towards satisfying the normalization condition, $\tilde{n}'(i) = 1$, everywhere.
This is done by formulating a loss function,
\begin{equation}
    \mathcal{L}(\theta) = \sum_{i \in \Omega} \mathcal{L}_{\theta}(i)
    \label{eq:L_theta_sum_in_Omega}
\end{equation}
where
%
\begin{equation}
    \mathcal{L}_{\theta}(i) = \left[ \ln \left( \sum_{j \in \mathcal{A}(i) \setminus \{ {\rm F}\}} \tilde{K}(i\to j) \, \frac{I(j; \theta)}{I(i; \theta)} \right) \right]^2.
    \label{eq:imp_func_log_loss}
\end{equation}

In systems exhibiting rare event problems, the optimal importance function, $I^{\textrm{opt}}(i)$, often yields extremely small values at microscopic states near state F.
Consequently, training a neural network to approximate the importance function can encounter numerical instabilities due to underflow problems.
%
%
We address this issue by optimizing the bias potential instead of the importance function.
Given Eq.~(\ref{eq:bias_pot}), even when the importance function takes extremely small values, the corresponding bias potential still takes values of manageable magnitude.
Thus parameterizing the bias potential $E_{\rm b}(i)$ using a neural network, $E_{\textrm{b}}(i;\,\theta)$, allows the neural network to operate within a numerically stable range, avoiding underflow problems.
%
%
%
Correspondingly, the loss function can be rewritten as,
%
\begin{equation}
    \mathcal{L}_{\theta}(i)
    = \left[ \ln \left\{ \sum_{j \in \mathcal{A}(i) \setminus \{ {\rm F}\}} \tilde{K}(i\to j) \, \exp \left( -\frac{E_{\textrm{b}}(j; \theta) - E_{\textrm{b}}(i; \theta)}{k_{\textrm{B}}T} \right) \right\} \right]^2.
    \label{eq:bias_pot_log_loss}
\end{equation}
This approach is analogous to a recently proposed method~\cite{mitchell2024}.

In practice, it is impossible to perform the sum in Eq.~(\ref{eq:L_theta_sum_in_Omega}) because the total number of microscopic states in high-dimensional systems is too large.
Fortunately, this is also unnecessary.  The normalization condition only needs to be well satisfied on microscopic states that lie along the transition paths (from F to S).
Hence in practice we replace the sum in Eq.~(\ref{eq:L_theta_sum_in_Omega}) with one over the microscopic states sampled by the importance sampling itself, using an adaptive sampling scheme.
%
%
%
%
%
The adaptive scheme initiates a random walk from state F, governed by the modified transition probabilities, $\tilde{K}'(i \rightarrow j)$.
During this data collection phase, the jump probabilities are determined by the current neural network representation of the bias potential, with its parameters held fixed.
Once a sufficient number of microscopic states have been accumulated, the neural network is updated by minimizing the loss function evaluated over the sampled microscopic states.
Afterwards, the algorithm returns to the data sampling stage with the updated bias potential.
This alternating cycle between sampling with fixed parameters and optimizing the network is repeated until the bias potential converges.
Furthermore, we apply a \textit{simulated annealing} protocol to the optimization of the bias potential.
The training process starts at an elevated temperature and then decreases in a stepwise manner to eventually reach the target temperature.
At each temperature, the bias potential is trained for a given number of epochs.
We found that this procedure helps avoid numerical instabilities and is beneficial for the optimization of the bias potential to converge.

A useful feature of our importance sampling framework is that a neural network trained on a coarse grid can be directly applied to accelerate transitions on a finer grid without retraining. 
This capability is particularly valuable because the length of transition paths increases as the grid resolution is refined, increasing the computational cost of bias-potential training, which requires adaptive path sampling.
Using an importance function learned on a computationally more tractable coarse grid, this approach avoid the heavy computational burden of optimizing bias potentials directly on the fine grid.
The trained bias potential can then be used to sample transition paths on the fine grid.
The branched random walk sampling method described in Section~\ref{sec:brw_method} brings the variance under control even when the bias potential is not the optimal bias potential.
\subsection{Sampling Initial Microscopic States}
\label{subsec:determine_initial_state}

The first step in the path sampling is to sample the microscopic state $i$ jumping from state F according to the transition probability $\tilde{K}({\rm F}\to i)$.
%
In low-dimensional systems, this can be accomplished by enumerating all the microscopic states $i$ accessible from F and then sampling them according to their transition probabilities.
In high-dimensional systems, it is not possible to enumerate all microscopic states $i$ accessible from F, and a computationally tractable approach is needed.

%
%

We achieve this by modifying the control functions, $f(i)$ and $s(i)$, from Eqs.~(\ref{eq:control_functions_f}) and (\ref{eq:control_functions_s}).
For instance, consider a modified control function $f(i)$ defined as follows.
\begin{equation}
    f(i) = \xi \, (\Delta x)^d \, \exp \left( -\frac{E(i) - E(\textrm{F})}{2k_{\textrm{B}}T} \right) \cdot \exp \left( -\frac{E_{\textrm{C}}(i)}{k_{\textrm{B}}T} \right) \cdot \mathbbm{1} \left( \| \textbf{x}(i) - \textbf{x}(\textrm{A}) \| < r_{\textrm{c}} \right),
    \label{eq:f_i_confine}
\end{equation}
where $\xi$ is a dimensionless constant parameter, $\textbf{x}(i)$ is the position vector of the state $i$ in $d$-dimensional system, $\|\cdot\|$ represents the Euclidean norm, 
and $E_{\textrm{C}}(i)$ is a newly introduced a confinement potential.
The confinement potential, $E_{\textrm{C}}(i)$, can be an arbitrary convex function having its global minimum at the microscopic state A.
For simplicity, the confinement potential can be chosen as,
\begin{equation}
    E_{\textrm{C}} (i) = \frac{1}{2\eta^2} \, \| \textbf{x}(i) - \textbf{x}(\textrm{A}) \|^2,
    \label{eq:confine_potential}
\end{equation}
where $\eta$ is a constant parameter.
Hence, the confinement potential smoothly ``confines'' the sampled states to be near the microscopic state A.
According to Eq.~(\ref{eq:transition_rate_extended}), the jump rate from state F to state $i$ is,
\begin{equation}
    r(\textrm{F} \rightarrow i) = \nu \, \xi \, (\Delta x)^d \, \exp \left( -\frac{E(i) - E(\textrm{F})}{k_{\textrm{B}}T} \right) \cdot \exp \left( -\frac{E_{\textrm{C}}(i)}{k_{\textrm{B}}T} \right) \cdot \mathbbm{1} \left( \| \textbf{x}(i) - \textbf{x}(\textrm{A} \| < r_{\textrm{c}} \right).
    \label{eq:r_F_i_confine}
\end{equation}

This formulation of the jump rate significantly simplifies the sampling of initial microscopic states $i$ from F, particularly in high-dimensional systems.
This is because Eq.~(\ref{eq:r_F_i_confine}) corresponds to the equilibrium distribution governed by the effective potential $E(i) + E_{\textrm{C}}(i) - E(\textrm{F})$.
Consequently, these states can be readily sampled using a standard Monte Carlo simulation, which remains computationally tractable even in high dimensions.
It is not necessary to compute the total jump rate out of state F because we exclude the time spent in F when calculating the transition rate (Section~\ref{sec:rate_estimate}).
Given the forward jump rate $r(\textrm{F} \to i)$, the reverse jump rate is determined by the requirement of detailed balance, i.e.,
%
\begin{equation}
    r(i \rightarrow \textrm{F}) = \nu \, \xi \, (\Delta x)^d \, \cdot \exp \left( -\frac{E_{\textrm{C}}(i)}{k_{\textrm{B}}T} \right) \cdot \mathbbm{1} \left( \| \textbf{x}(i) - \textbf{x}(\textrm{A}) \| < r_{\textrm{c}} \right) \, .
    \label{eq:r_i_F_confine}
\end{equation}
%
%
Notice that the jump rate from $i$ to F becomes greater as the microscopic state $i$ gets closer to microscopic state A, provided that the confinement potential is a convex function with the minimum at A.

\section{Results}
\label{sec:results}
The importance sampling framework integrated with neural network, is applied to 2-dimensional and 14-dimensional systems to demonstrate its effectiveness and scalability.

\subsection{2-Dimensional System}
\label{sec:2-dim}

A 2-dimensional potential energy landscape (in units of eV) described in Section~\ref{sec:methods} and shown in Figure~\ref{fig:pel_discrete} can be analytically written as follows.
%
\begin{equation}
    E_{\textrm{2D}} (x_1, \, x_2) = 0.02 \, x_2 + \frac{1}{6} \left[ 4 \, (1-x_1^2-x_2^2)^2 + 2(x_1^2-2)^2 + \left\{(x_1+x_2)^2-1\right\}^2 + \left\{(x_1-x_2)^2-1\right\}^2 - 2 \right],
    \label{eq:pel_2d_again}
\end{equation}
The landscape is discretized over the domain $x_1, \, x_2  \in \left[-1.5, \, 1.5\right]$ using equal grid spacing $\Delta x_1 = 0.1$ and $\Delta x_2 = 0.1$.
The energy minima are located at microscopic state A $(-1.1, \, 0.0)$ and microscopic state B $( 1.1, \, 0.0)$, which correspond to two metastable states that contain them.
The two saddle points are located at S$_1$ $(0.0, \, 1.0)$ and S$_2$ $(0.0, \, -1.0)$, and their energy barriers are $1.02 \, \rm eV$ at $\rm S_1$ and $0.98 \, \rm eV$ at $\rm S_2$.
Transitions from A to B are considered as rare events, particularly at low temperatures.
The primary transition pathways from A to B proceed through regions proximal to the saddle points S$_1$ and S$_2$.
The functions $f(i)$ and $s(i)$ to control the jump between microscopic state $i$ and F or S follow Eqs.~(\ref{eq:control_functions_f}-\ref{eq:control_functions_s}) and the parameters of the truncated Gaussians used in this example are, $\epsilon = 0.1, \, \sigma = 5 \times 10^{-3}, \, r_{\textrm{c}} = 0.3$, $E(\textrm{F}) = -0.5 \, \textrm{eV}$, $E(\textrm{S}) = -0.5 \, \textrm{eV}$.

\begin{figure*}[htbp!]
    \centering
    \subfloat[]{%
        \includegraphics[width=.34\textwidth]{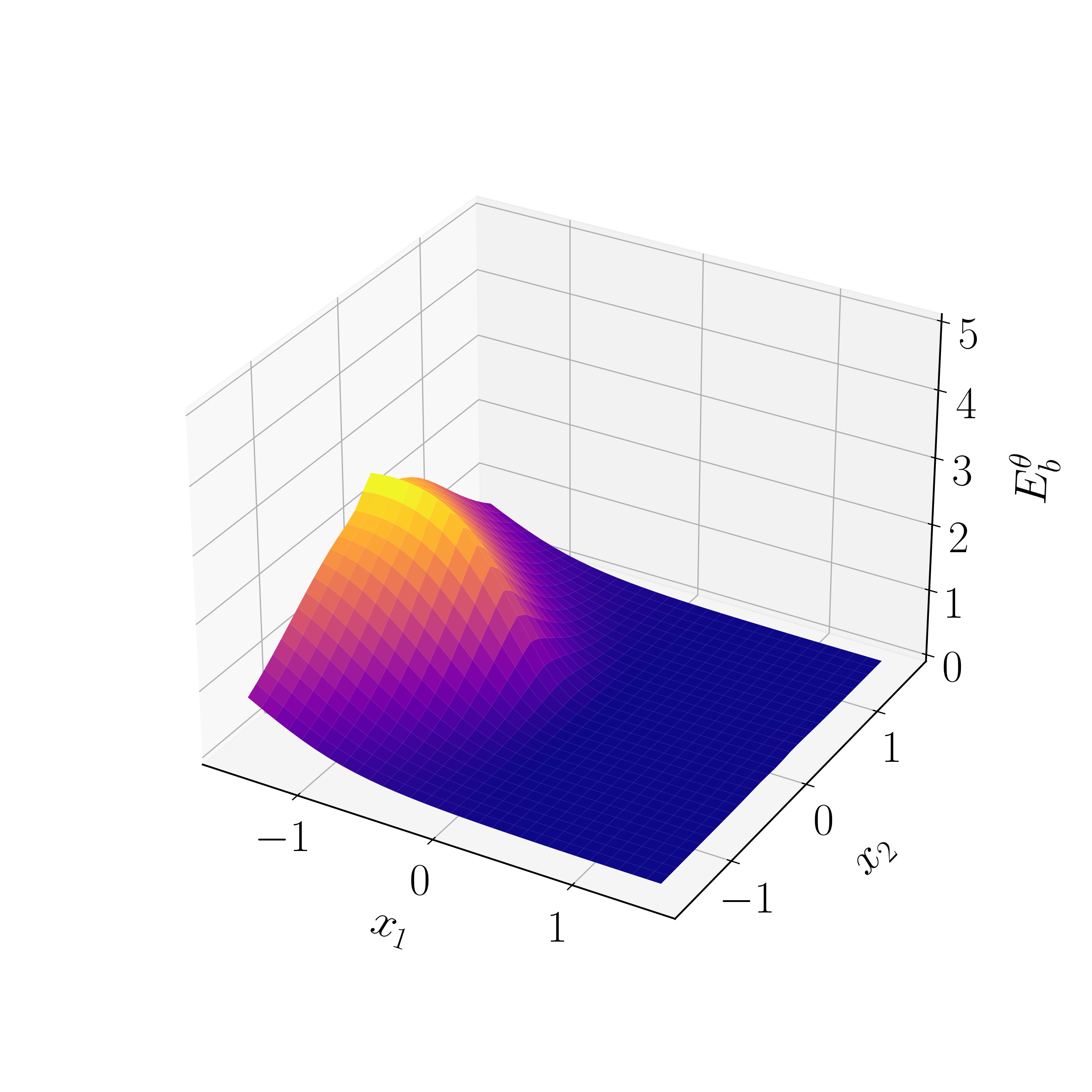}%
        \label{subfig:eb_nn_30e_2d}%
    }
    \subfloat[]{%
        \includegraphics[width=.28\textwidth]{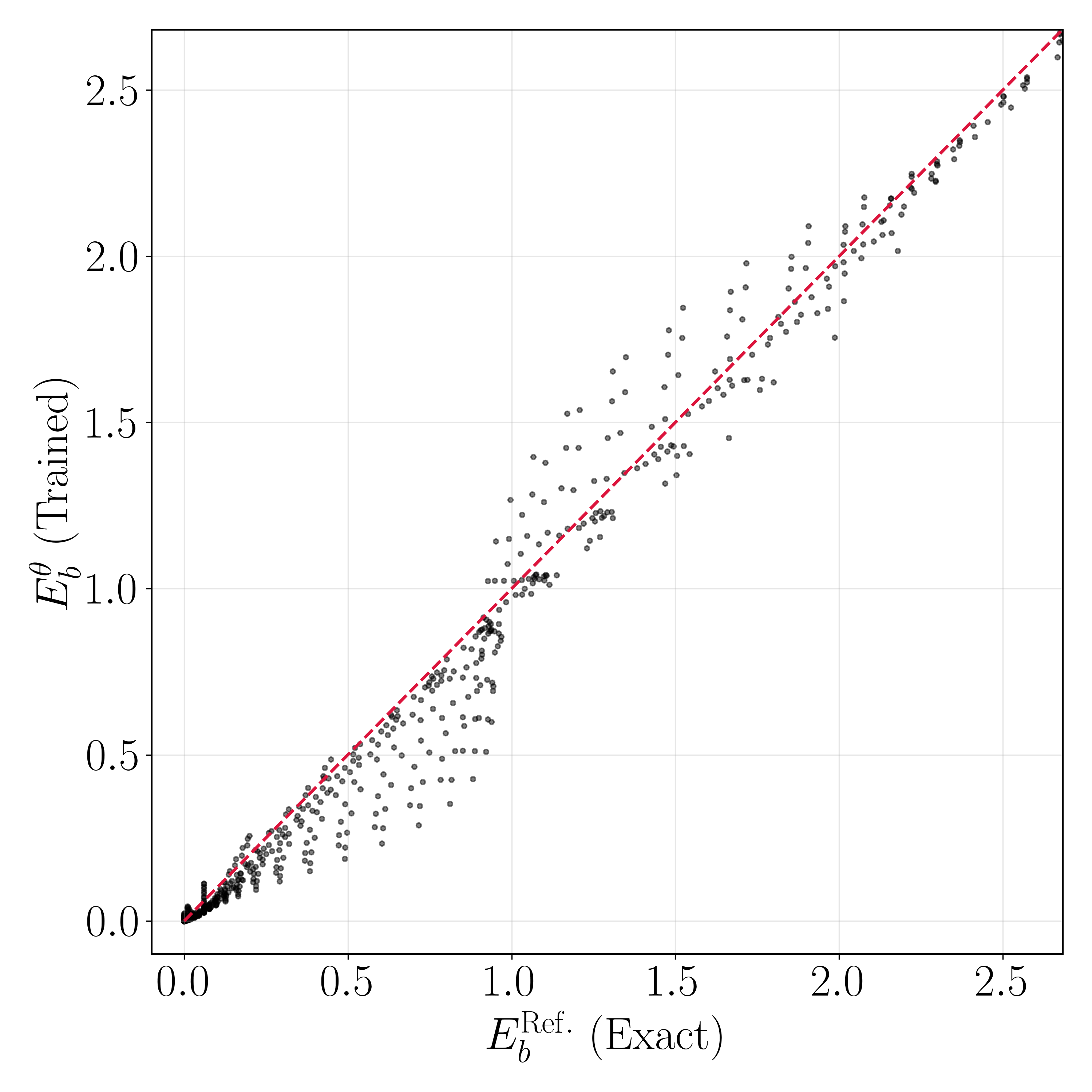}%
        \label{subfig:parity_eb_nn_30e_2d}%
    }
    \subfloat[]{%
        \includegraphics[width=.36\textwidth]{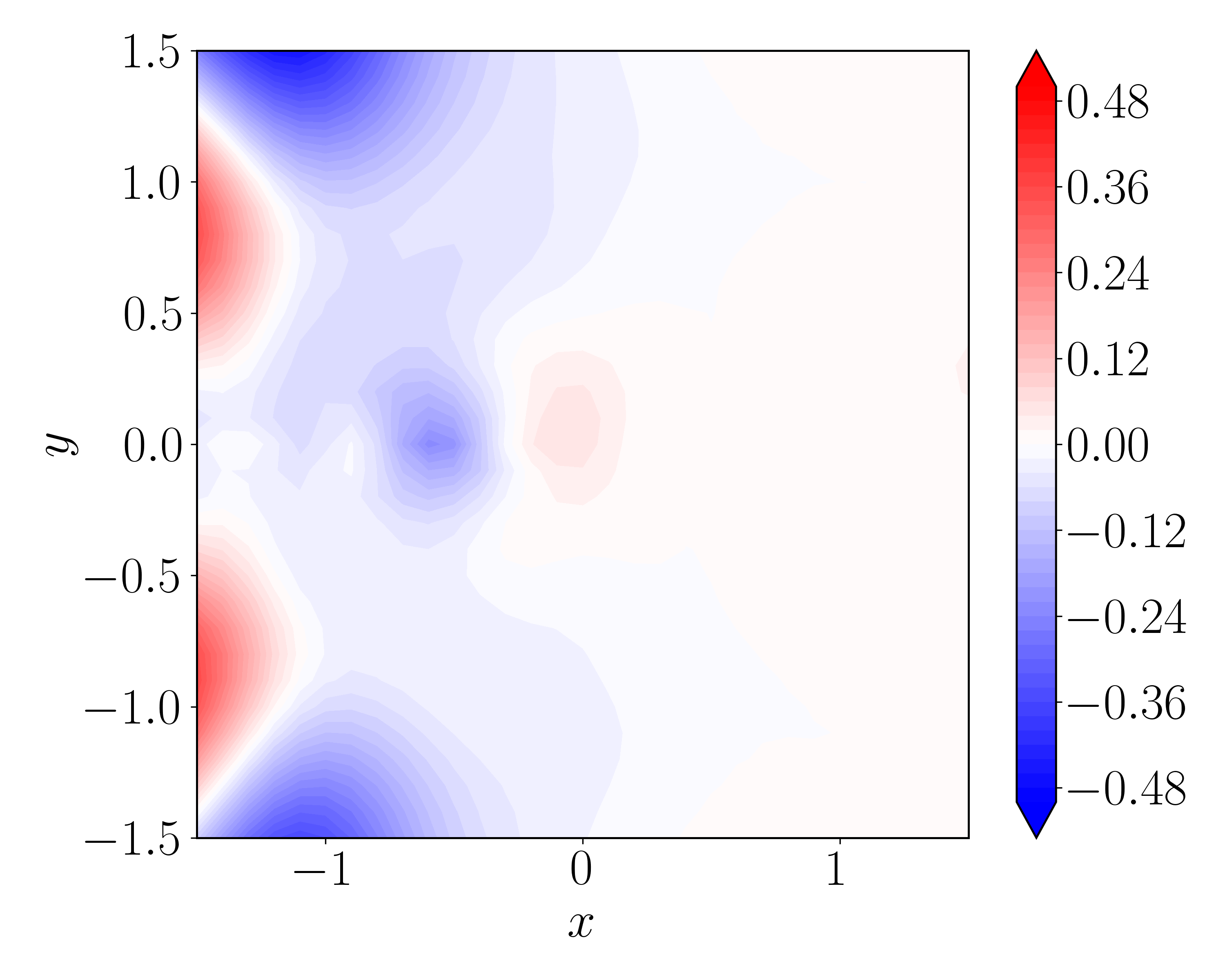}%
        \label{subfig:error_eb_nn_30e_2d}%
    }\\
    \subfloat[]{%
        \includegraphics[width=.34\textwidth]{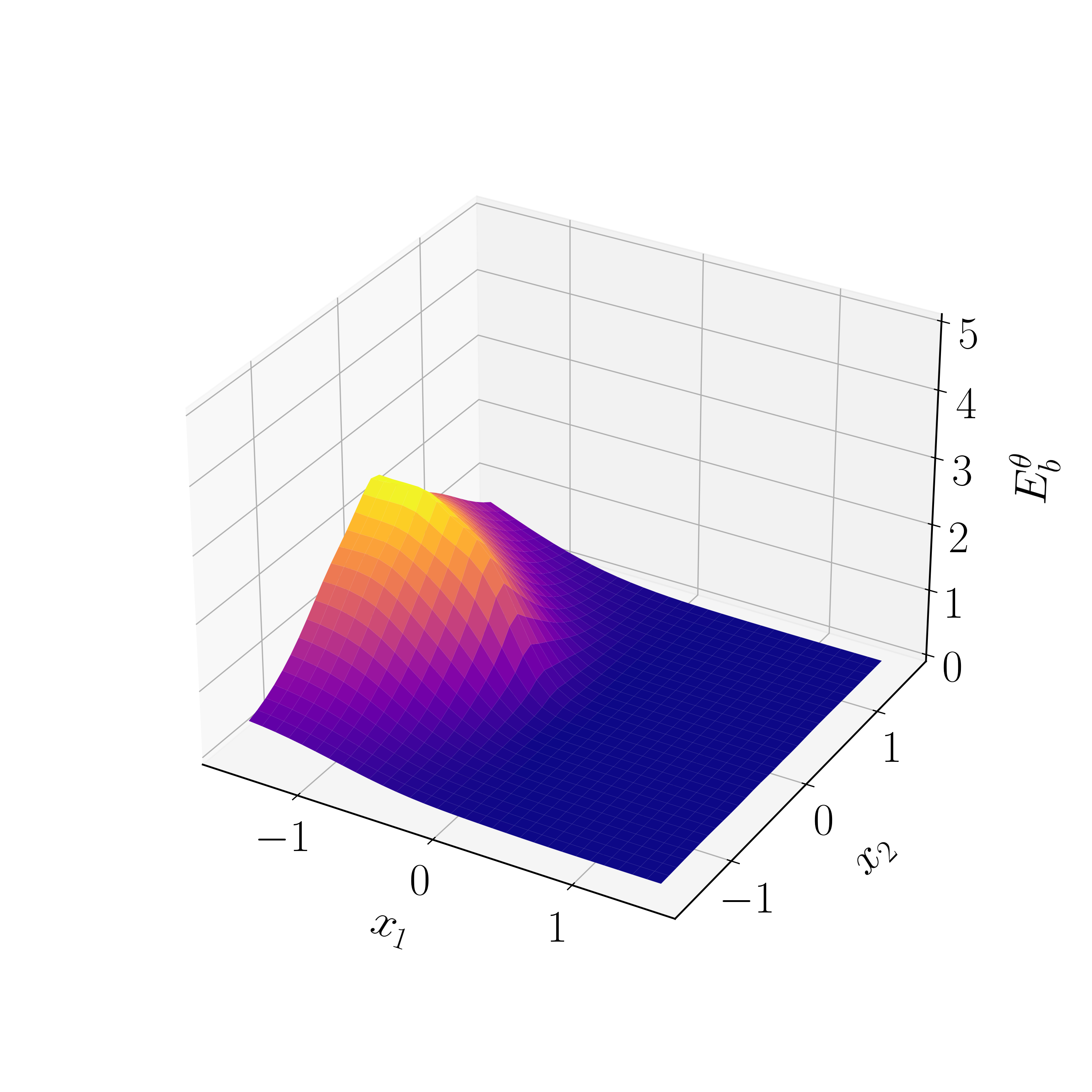}%
        \label{subfig:eb_nn_100e_2d}%
    }
    \subfloat[]{%
        \includegraphics[width=.28\textwidth]{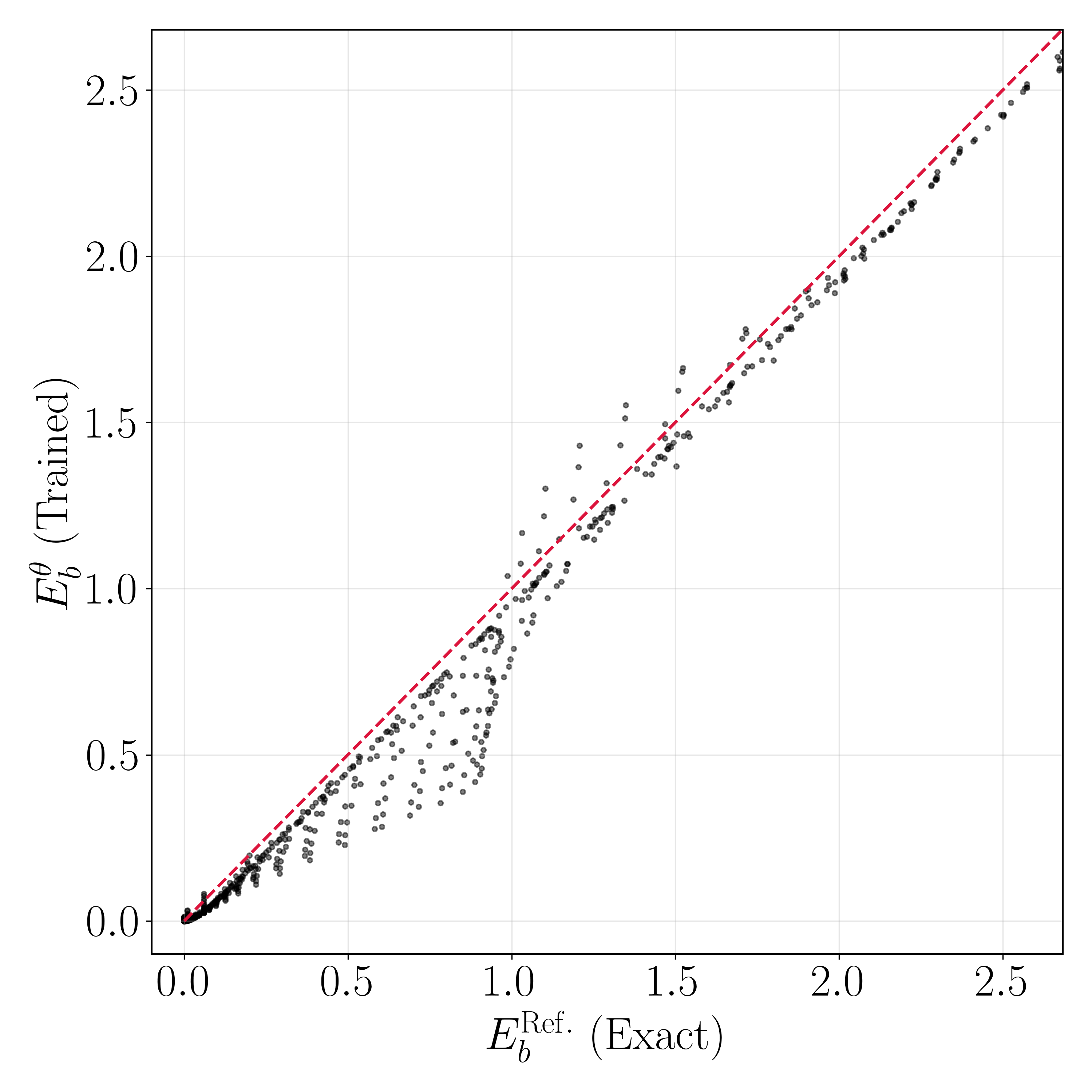}%
        \label{subfig:parity_eb_nn_100e_2d}%
    }
    \subfloat[]{%
        \includegraphics[width=.36\textwidth]{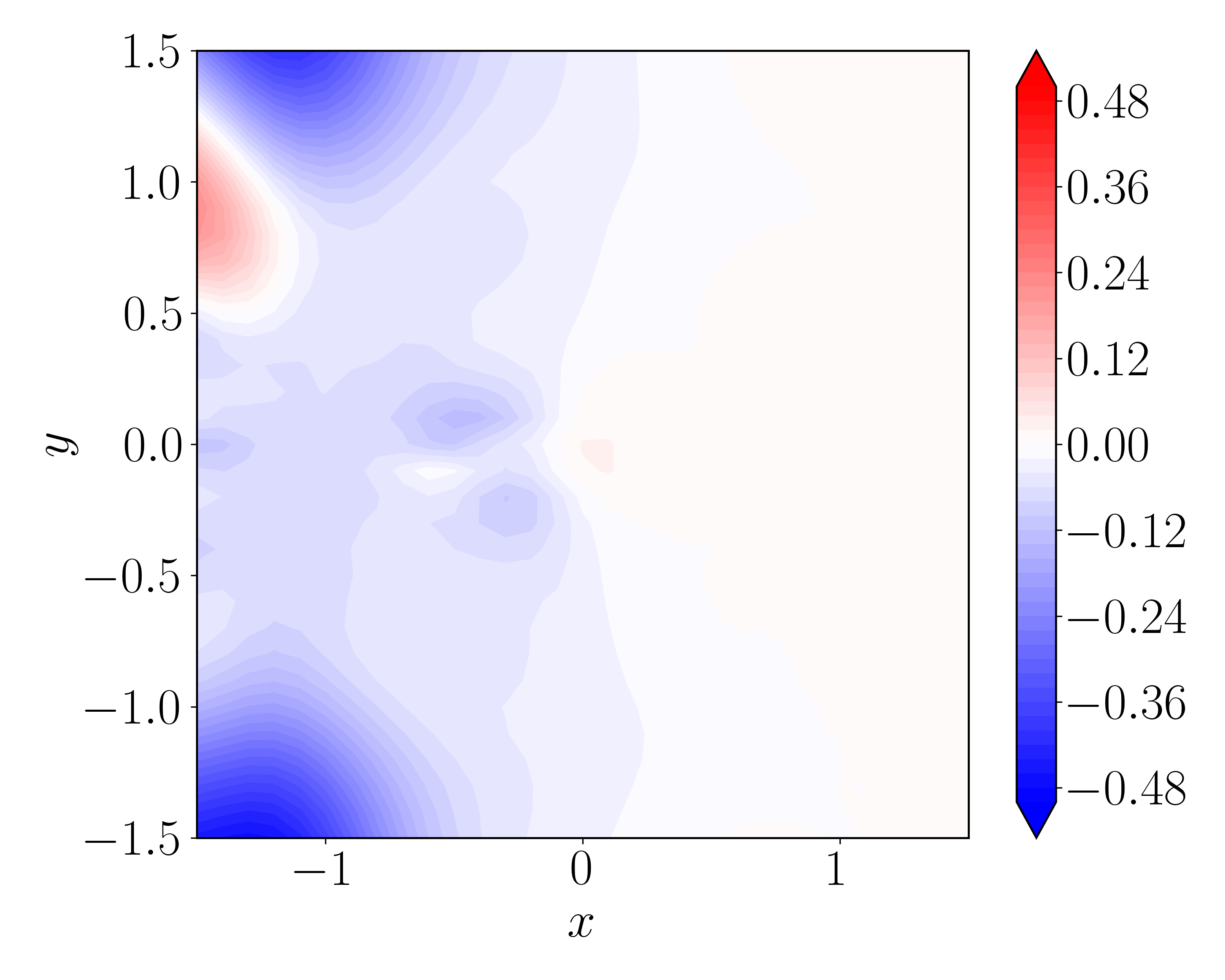}%
        \label{subfig:error_eb_nn_100e_2d}%
    }
    \caption{Neural network bias potential over a 2-dimensional domain after (a) training for 30 epochs and (d) 100 epochs (at each temperature). (b) Parity plots between the neural network bias potential and the exact optimal bias potential after training for (b) 30 epochs and (e) 100 epochs (at each temperature).
    Differences between the neural network bias potential and the exact solution after training for (c) 30 epochs and (f) 100 epocs (at each temperature).}
    \label{fig:2d_bias}
\end{figure*}
The neural network architecture comprises two hidden layers, each containing 30 perceptrons with \texttt{tanh} activation functions.
%
%
%
%
We employ a simulated annealing protocol where the training initially starts at $5800$~K and subsequently cooled to the target of $500$~K over a sequence of $8$ discrete stages.
%
%
At each temperature, the training is performed for $30$ or $100$ epochs using the ADAM~\cite{kingma2014adam} algorithm with a learning rate of $10^{-3}$.
%
%
%
%
%
For the 2-dimensional system, since we can enumerate all the states and store the entire transition probability matrix, $K(i \rightarrow j)$, the optimal importance function, $I^{\textrm{opt}}(i)$, can be solved numerically~\cite{Cai2002}, and the optimal bias potential can be obtained from Eq.~(\ref{eq:bias_pot}).
Figure~\ref{subfig:eb_nn_30e_2d} plots the neural network bias potential trained for 30 epochs (at each temperature) and Figure~\ref{subfig:parity_eb_nn_30e_2d} shows a parity plot comparing the neural network bias potential  with the exact optimal bias potential~\cite{Cai2002}.
Similarly, Figure~\ref{subfig:eb_nn_100e_2d} plots the neural network bias potential trained for 100 epochs (at each temperature), and Figure~\ref{subfig:parity_eb_nn_100e_2d} compares it with the exact optimal bias potential.
%
%
The neural network bias potential trained after 100 epochs  (at each temperature) shows a substantially improved agreement with the exact optimal bias potential, compared to the one trained after 30 epochs (at each temperature).
Notably, the largest deviations (i.e., error) between the neural network bias potential and the exact optimal bias potential occur in sparsely sampled microscopic states that lie far from the dominant transition pathways, as shown in Figure~\ref{subfig:error_eb_nn_30e_2d} and \ref{subfig:error_eb_nn_100e_2d}.
%

\subsection{Success Probability Estimate}

We demonstrate in the 2-dimensional system that the success probability estimator given by Eq.~(\ref{eq:PS_F_by_imp_sample_R_W}) is unbiased even when the importance function is not exactly the same as the optimal solution.
%
%
The optimal importance function, $I^{\textrm{opt}}(i)$, can be used to compute the reference success probability, $p_{\textrm{S}}(\textrm{F})$, as follows,
%
%
\begin{equation}
    p_{\textrm{S}}(\textrm{F}) = \frac{I^{\textrm{opt}}(\textrm{F})}{I^{\textrm{opt}}(\textrm{S})}.
    \label{eq:p_S_F_I_opt}
\end{equation}
The reference success probability is $2.1899 \times 10^{-13}$ at $T = 500~\textrm{K}$ ($k_{\textrm{B}}T = 0.043087 \, \textrm{eV}$).

%
The neural network bias potential, $E_{\textrm{b}}(i;\,\theta)$, is employed to accelerate the transition of the system and sample 100 successful paths.
The scores of these sampled successful paths were then used to estimate the success probability, $p^{\theta}_{\textrm{S}}(\textrm{F})$, via Eq.~(\ref{eq:PS_F_by_imp_sample_R_W}).
This accelerated sampling was repeated 100 times to improve statistical accuracy of the success probability estimator.

\begin{figure*}[htbp!]
    \subfloat[]{%
        \includegraphics[width=.50\linewidth]{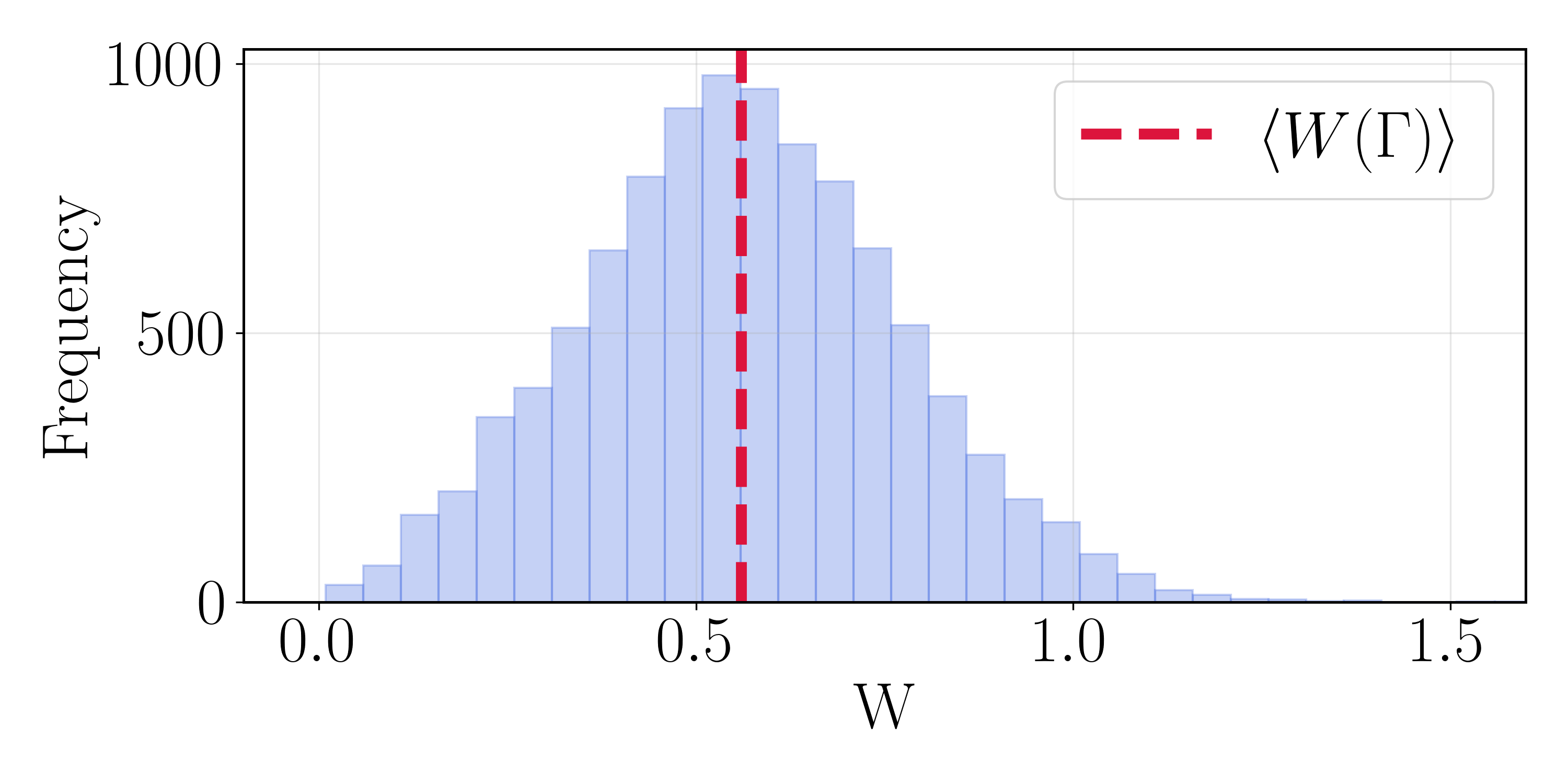}%
        \label{subfig:w_nn_30e_31n_2d}%
    }\hfill
    \subfloat[]{%
        \includegraphics[width=.50\linewidth]{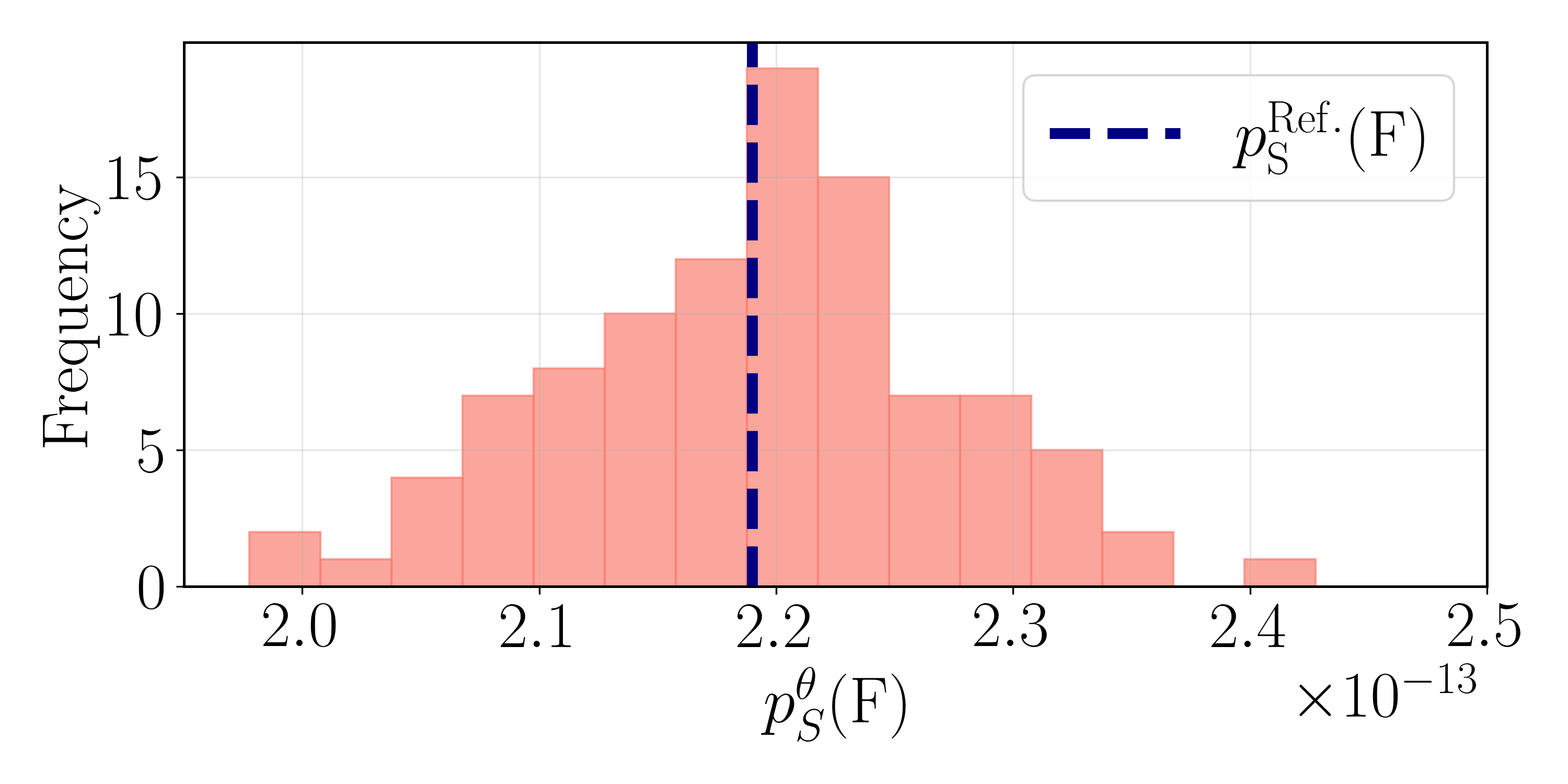}%
        \label{subfig:ps_nn_30e_31n_2d}%
    }\\
    \subfloat[]{%
        \includegraphics[width=.50\linewidth]{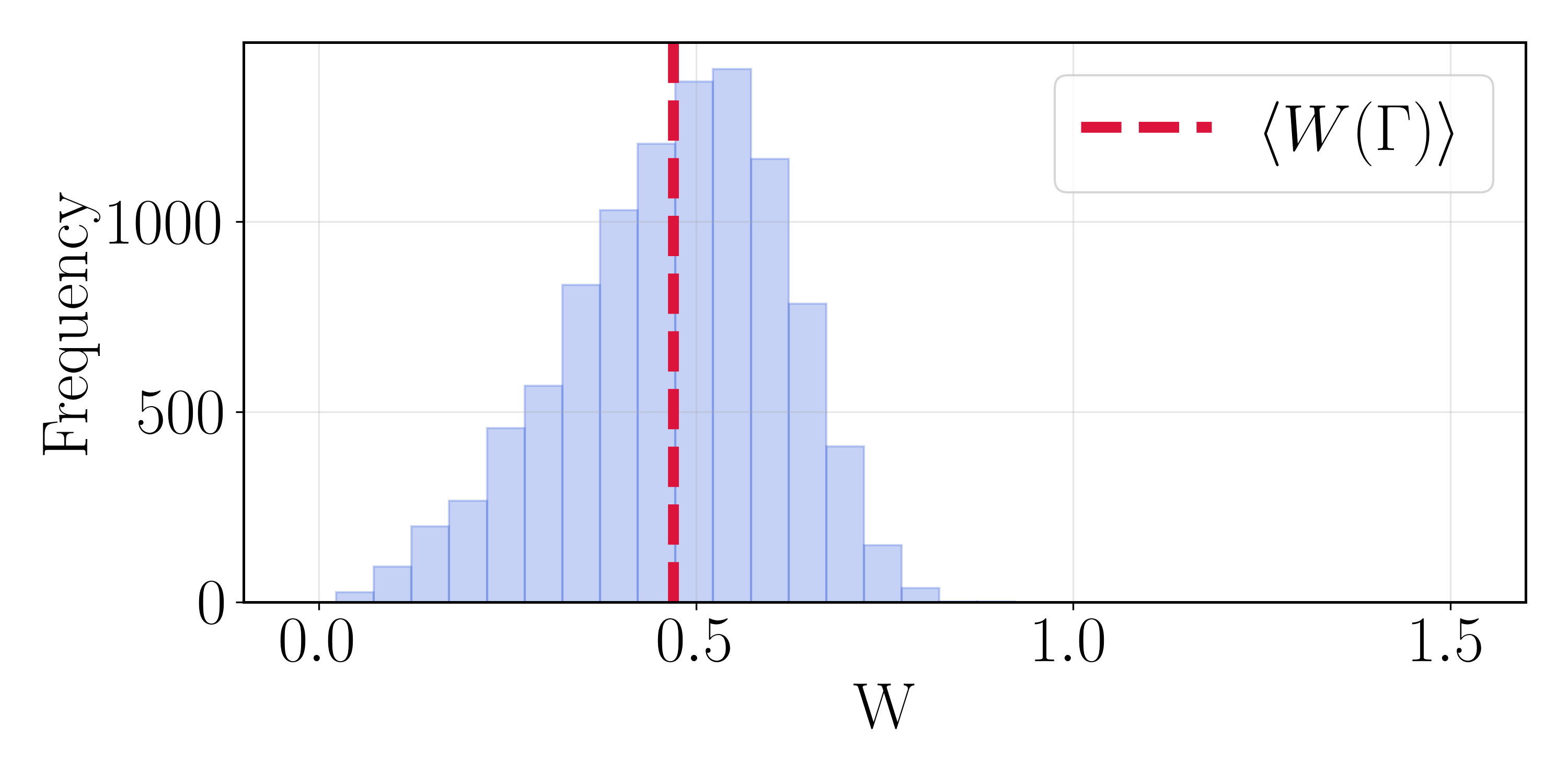}%
        \label{subfig:w_nn_100e_31n_2d}%
    }\hfill
    \subfloat[]{%
        \includegraphics[width=.50\linewidth]{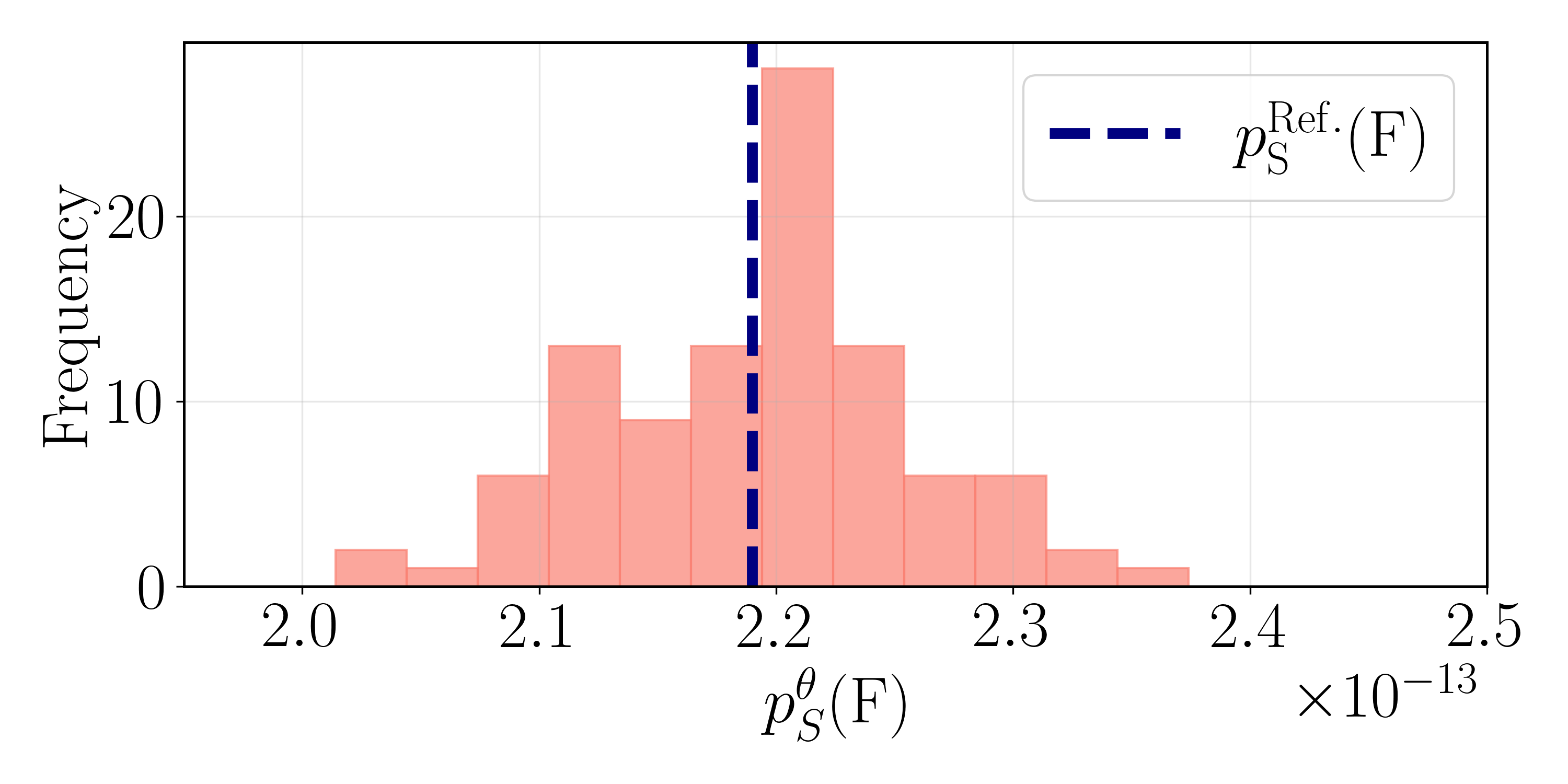}%
        \label{subfig:ps_nn_100e_31n_2d}%
    }
    \caption{Weight distribution of successful paths for importance sampled success paths using neural network bias potentials trained for (a) 30 epochs and (c) 100 epochs (at each temperature).
    Distribution of success probability estimates for bias potentials trained for (b) 30 epochs and (d) 100 epochs (at each temperature).
    }
    \label{fig:succ_prob_estim}
\end{figure*}
Figures~\ref{subfig:w_nn_30e_31n_2d} and \ref{subfig:w_nn_100e_31n_2d} present the weight distributions of the successful paths using models trained for 30 epochs and 100 epochs at each temperature, respectively.
The deviation of the mean weights from unity indicates that the neural network importance functions are not optimal.
Figure~\ref{subfig:ps_nn_30e_31n_2d} and Figure~\ref{subfig:ps_nn_100e_31n_2d} show the corresponding distributions of success probability estimates.
The estimated success probabilities are $2.1921 \pm 0.0082 \times 10^{-13}$ and $2.1937 \pm 0.0065 \times 10^{-13}$ for models trained for 30 epochs and 100 epochs (at each temperature), respectively, which are close to the reference success probability.
While the success probability estimates exhibit some variance, their mean values provide unbiased estimates of the true success probability, $p_{\textrm{S}}(\textrm{F})$.
Notice that the neural network importance function trained for a longer duration (i.e., 100 epochs at each temperature) exhibits sharper weight and success probability distributions, indicating an improved approximation of the optimal importance function.
\subsection{Branching Random Walk}
\begin{figure*}[htbp!]
    \subfloat[]{%
        \includegraphics[width=.50\linewidth]{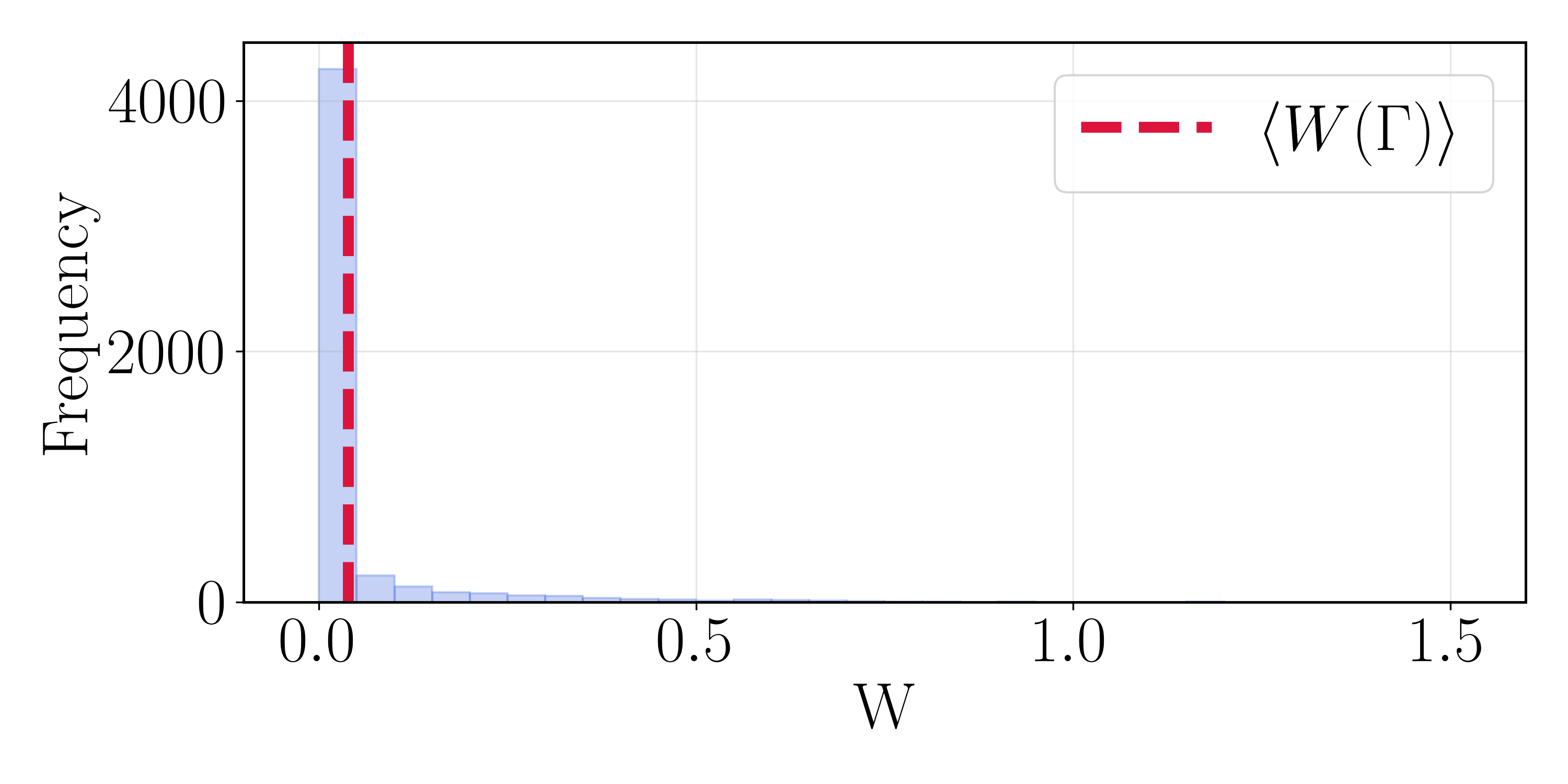}%
        \label{subfig:w_nn_30e_121n_2d}%
    }\hfill
    \subfloat[]{%
        \includegraphics[width=.50\linewidth]{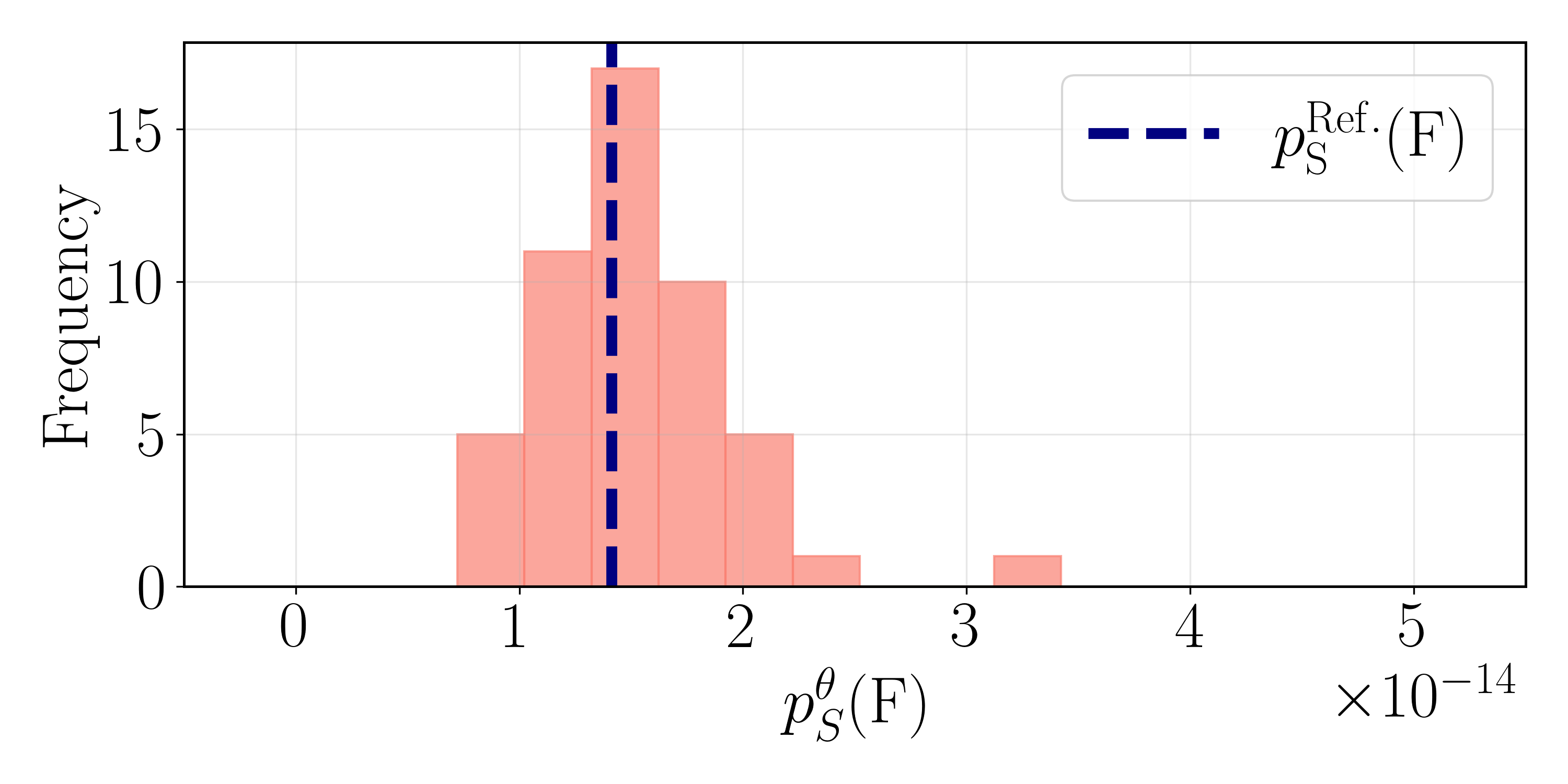}%
        \label{subfig:ps_nn_30e_121n_2d}%
    }\\
    \subfloat[]{%
        \includegraphics[width=.50\linewidth]{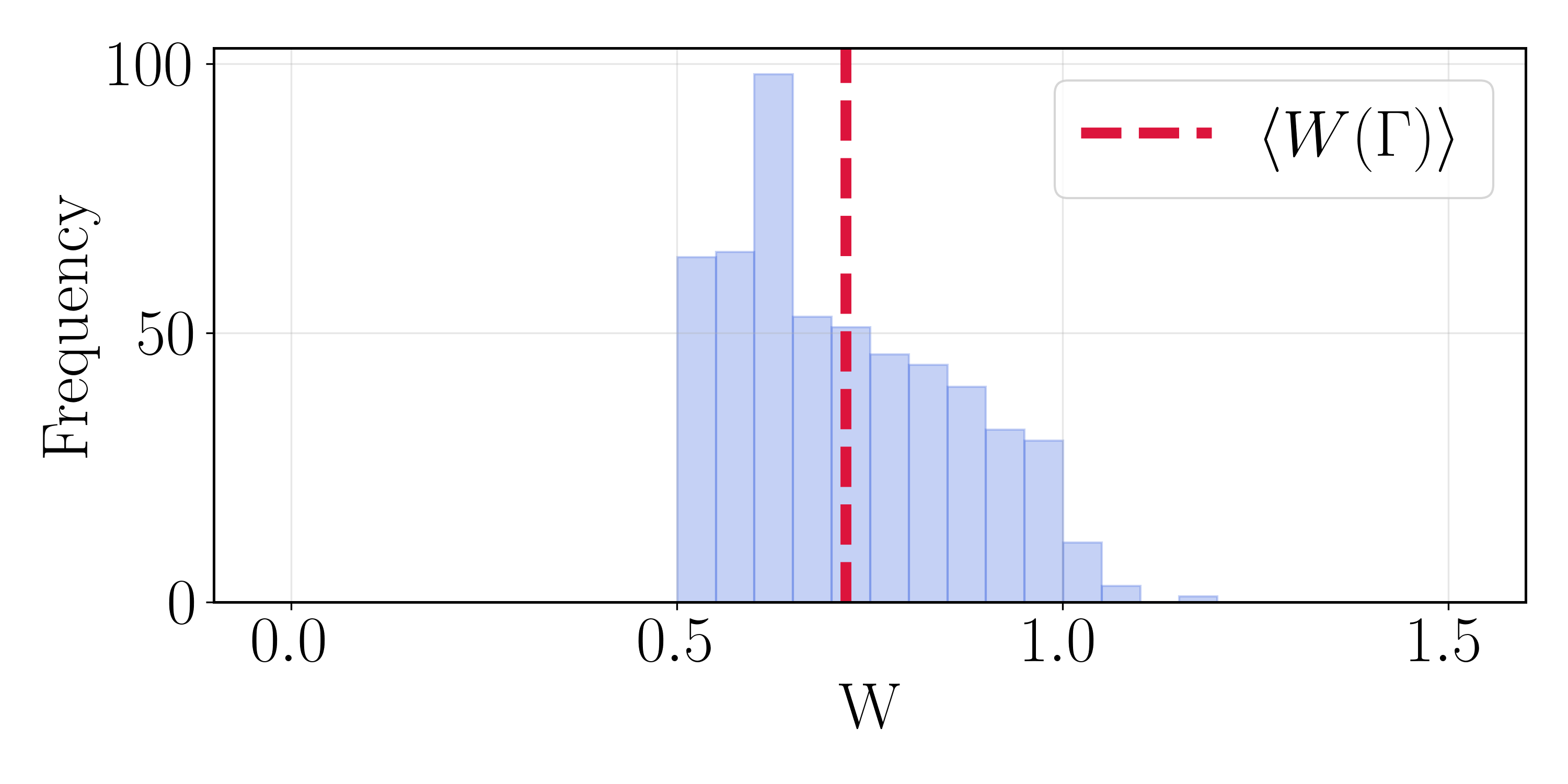}%
        \label{subfig:w_nn_30e_121n_2d_brw_0.5_1.2}%
    }\hfill
    \subfloat[]{%
        \includegraphics[width=.50\linewidth]{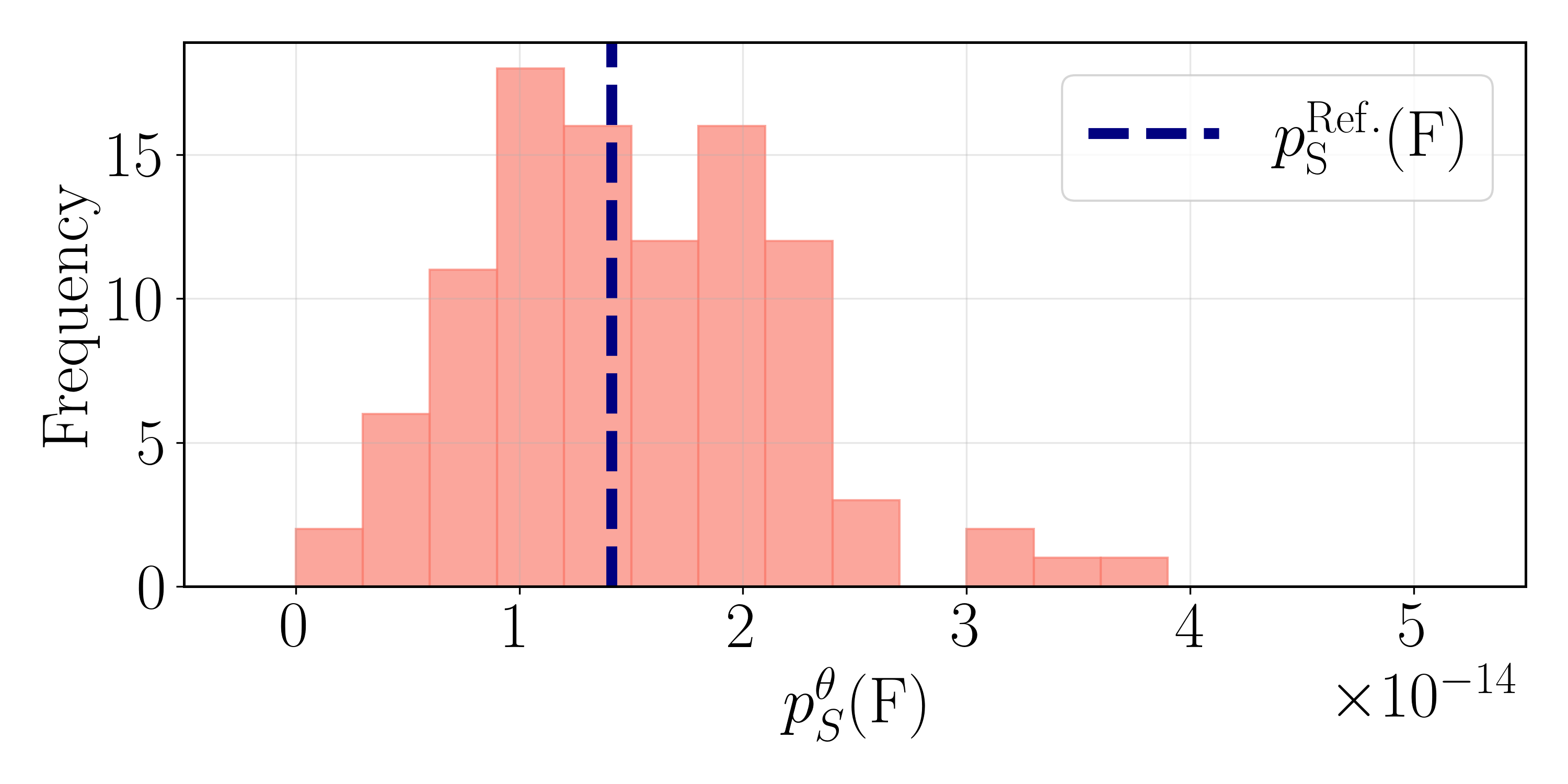}%
        \label{subfig:ps_nn_30e_121n_2d_brw_0.5_1.2}%
    }
    \caption{Weight distributions of the sampled paths (a) without and (c) with the BRW process. Distribution of success probability estimates (b) without and (d) with the BRW process.}
    \label{fig:succ_prob_estim_brw}
\end{figure*}
\noindent To study the robustness of the proposed method against approximation errors, we intentionally introduce a discretization mismatch between the training and sampling phases. 
Specifically, the neural network importance function trained on a coarse grid ($\Delta x_1 = \Delta x_2 = 0.1$) is employed to bias the dynamics of the system on a refined grid ($\Delta x_1 = \Delta x_2 = 0.025$).
Figure~\ref{subfig:w_nn_30e_121n_2d} shows the weight distribution of successful paths sampled under this condition.
%
%
Sampling was performed without the branching random walk (BRW) process over 50 batchs, with 100 paths sampled per batch.
%
%
The majority of the sampled successful paths exhibit near-zero weights, meaning that these are unimportant paths with minimal contribution to the overall estimate.
Figure~\ref{subfig:ps_nn_30e_121n_2d} shows the distribution of the success probability estimates.
The estimated success probability is $1.5198 \pm 0.0627 \times 10^{-14}$, while the reference (exact) success probability is $1.4120 \times 10^{-14}$.

Using the branching random walk (BRW) process described in Section~\ref{sec:brw_method}, the weights of successful paths can be maintained within a specified range.
By controlling the weight distribution, paths with sufficiently small weights are terminated early, thereby enhancing the efficiency of the sampling process.
Sampling was done with the BRW process over 100 batches, with 100 paths sampled per batch to match the variance of the success probability estimates obtained from sampling without the BRW process.
%
%
Figure~\ref{subfig:w_nn_30e_121n_2d_brw_0.5_1.2} shows the distribution of weights using the BRW process, controlled to be within $[0.5, \, 1.2]$. 
Figure~\ref{subfig:ps_nn_30e_121n_2d_brw_0.5_1.2} shows the distribution of the success probability estimates using the BRW process. 
The estimated success probability using the BRW process is $1.5210 \pm 0.0690 \times 10^{-14}$.

While sampling with and without the BRW process both yield unbiased estimates of the success probability, the BRW process exhibits superior computational efficiency compared to sampling without the BRW process.
The total number of Monte Carlo steps without the BRW process is 7,967,789 steps, whereas the total number with the BRW process is 958,928 steps.
This corresponds to an approximately 8-fold speedup.
These results show that using the BRW process allows us to achieve similar variance levels in the success probability estimate at a reduced computational cost.
\subsection{Transition rate and mechanisms}
\label{subsec:transition_rate_mechanisms}

\begin{figure*}[htbp!]
    \centering
    \subfloat[]{%
        \includegraphics[width=.33\textwidth]{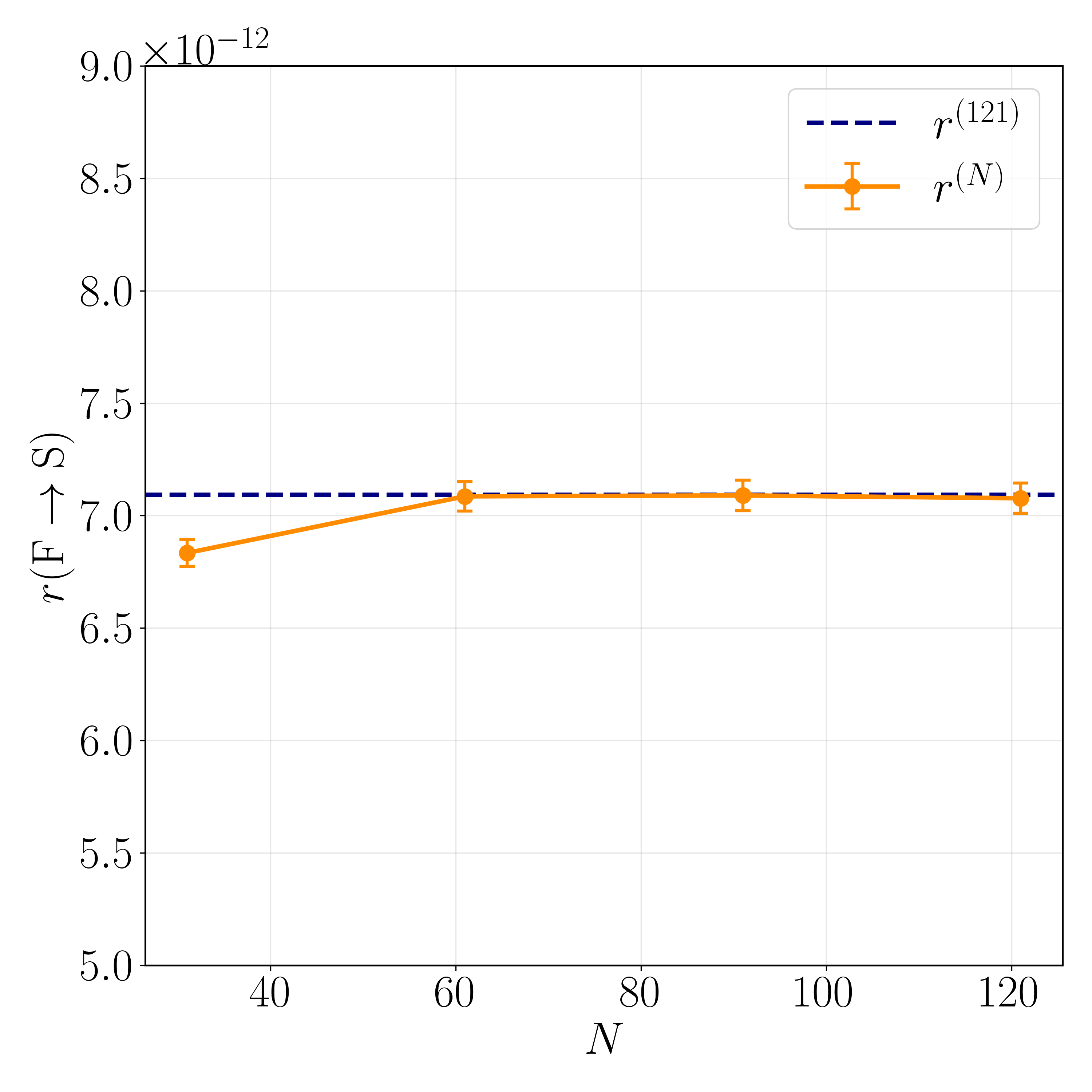}%
        \label{subfig:rate_nn_conv_annh}%
    }
    \subfloat[]{%
        \includegraphics[width=.32\textwidth]{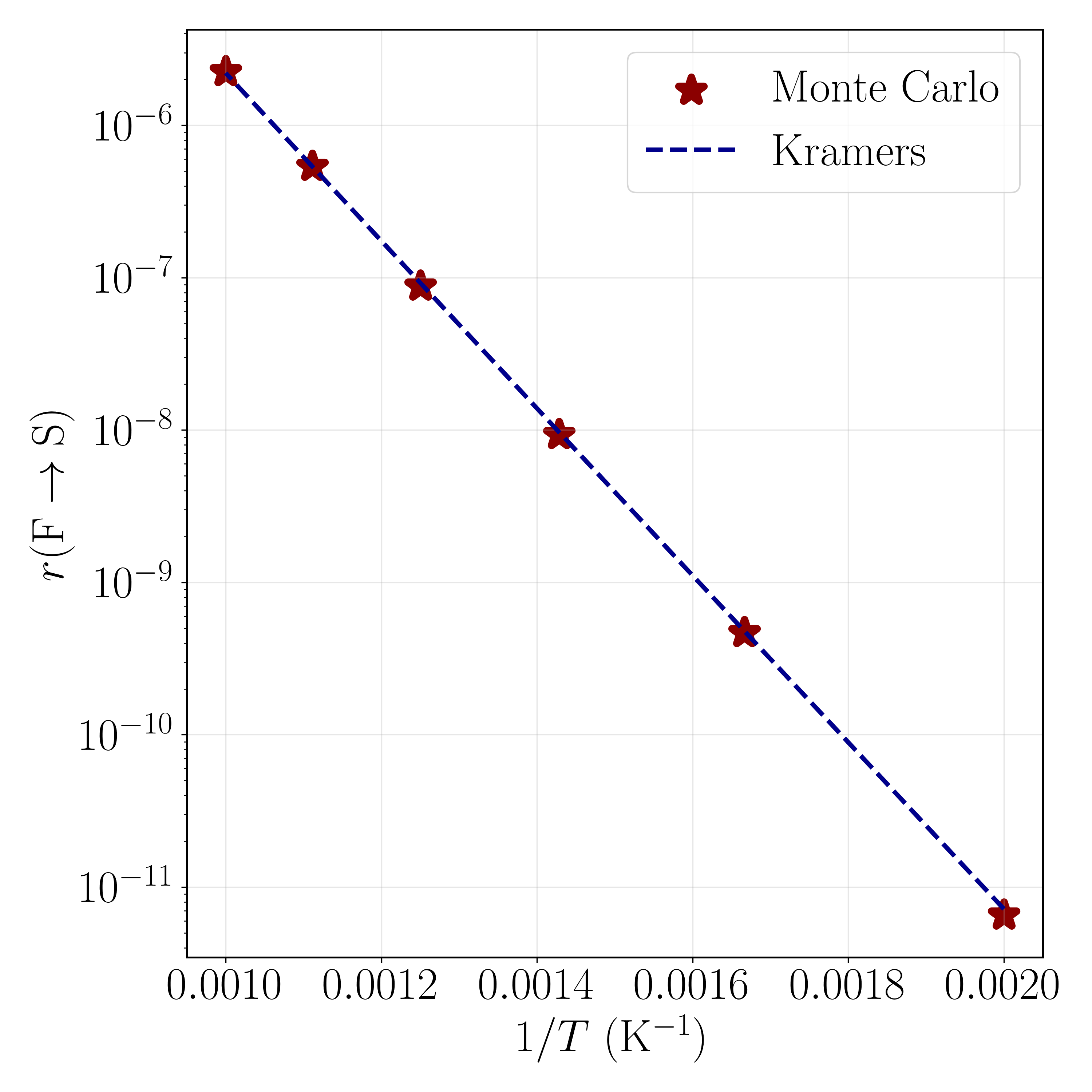}%
        \label{subfig:rate_nn_arrh_annh}%
    }
    \subfloat[]{%
        \includegraphics[width=.32\textwidth]{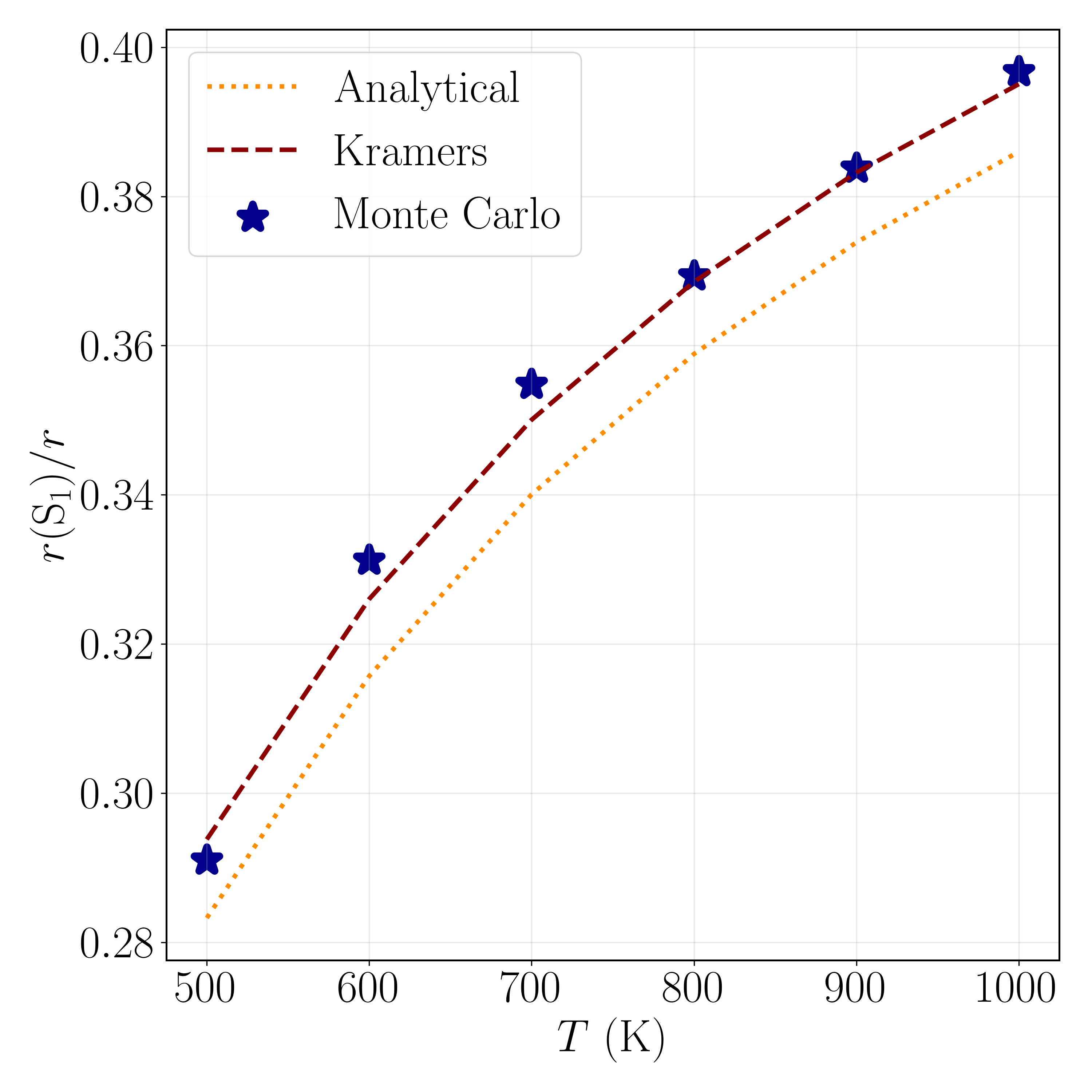}%
        \label{subfig:rate_nn_frac_annh}%
    }
    \caption{(a) Transition rate at temperature $T=500$~K computed using 
    importance sampling with BRW
    for systems with different grid sizes.
    The dotted line represents the transition rate with grid size $\Delta x = 0.025$.
    %
    (b) Temperature dependence of the transition rate computed (stars) at fixed grid spacing $\Delta x = 0.025$.
    The theoretical prediction from Kramers' rate theory is shown as dashed line for comparison.
    (c) Fraction of the transitions going through the top channel near saddle point $S_1$ as a function of temperature.
    The Monte Carlo sampling results are shown as stars.
    %
    Theoretical predictions based on barrier energy differences only (dotted line) and Kramers' theory (dashed line) are provided for comparison.
    %
    %
    }
    \label{fig:tran_rate}
\end{figure*}
\noindent We demonstrate that the transition rates estimated using the method described in Section~\ref{sec:rate_estimate} are consistent across various grid sizes.
Figure~\ref{subfig:rate_nn_conv_annh} shows transition rates estimated for systems with different number of grid points in each dimension (corresponding to difference grid sizes $\Delta x = 3.0/(N-1)$) at $T=500$~K.
The blue dashed line indicates the exact transition rate of $6.9191 \times 10^{-12}$ for the 2-dimensional system with grid size of $\Delta x = 0.025$.
%
%
The neural network bias potential trained on a system with grid size $\Delta x = 0.1$ 
was used to estimate the success probability for all systems.
The estimated transition rates across systems with varying grid sizes converge with increasing $N$, and to a value in good agreement with the exact transition rate.
The error bar corresponds to one standard deviation from the mean of the transition rate estimates.
%

In order to validate the accuracy of the estimated transition rates, we compare them with Kramers rate theory~\cite{Kramers1940, eyring1935} for overdamped dynamics at various temperatures.
Figure~\ref{subfig:rate_nn_arrh_annh} shows an Arrhenius plot of the estimated transition rates, demonstrating good agreement with Kramers' rate theory.
The transition rate expression from Kramers' rate theory reads,
\begin{align}
    r(\textrm{A} \rightarrow \textrm{S}_1 \rightarrow \textrm{B}) &= \nu_1 \, \exp \left( -\frac{E(\textrm{S}_1) - E(\textrm{A})}{k_{\textrm{B}}T} \right),
    \label{eq:Erying_Kramers_r1}\\
    r(\textrm{A} \rightarrow \textrm{S}_2 \rightarrow \textrm{B}) &= \nu_2 \, \exp \left( -\frac{E(\textrm{S}_2) - E(\textrm{A})}{k_{\textrm{B}}T} \right),
    \label{eq:Erying_Kramers_r2}
\end{align}
where $\nu_1$ and $\nu_2$ are frequency prefactors related with the saddle points, S$_1$ and S$_2$, respectively.
More details on the application of Kramers rate theory to the model here is given in Appendix~\ref{appdx:eyring_kramers_theory}.
%

%
We further demonstrate that the accelerated sampling using the neural network bias potential accurately resolves the transition rates associated with distinct reaction channels.
Figure~\ref{subfig:rate_nn_frac_annh} presents the fraction of the transition rate that proceeds through the saddle point S$_1$ at different temperatures.
At $T=500~\textrm{K}$, the accelerated Monte Carlo simulation estimates this fraction to be $0.2910$, while at $T=1000~\textrm{K}$, the fraction increases to $0.3967$.
These results show good agreement with Kramers rate theory predictions of 0.2938 and 0.3951 at $T=500~\textrm{K}$ and $T=1000~\textrm{K}$, respectively (dashed line).
Notably, the fraction of transitions through S$_1$ cannot be determined solely from the energy difference between the two saddle points, i.e. by assuming $\nu_1 = \nu_2$ in Eqs.~(\ref{eq:Erying_Kramers_r1})-(\ref{eq:Erying_Kramers_r2}).
Had we made that assumption, the prediction of the simplying analytic theory would be the dotted line in Fig.~\ref{subfig:rate_nn_frac_annh}, which significantly deviates from the Monte Carlo results.
%
%
The observed temperature dependence arises from distinct rate pre-factors $\nu_1$ and $\nu_2$ associated with the different local curvatures of the potential energy landscapes at the saddle points S$_1$ and S$_2$. 
%
\subsection{14-dimensional system}
\label{sec:14-dim}

A 14-dimensional potential energy landscape (in units of eV) can be constructed by extending the 2-dimensional system described in Eq.~(\ref{eq:pel_2d_again}) through the addition of harmonic terms in the remaining coordinates.
\begin{equation}
    E_{\rm 14D}(x_1, x_2, ..., x_{14}) = E_{\rm 2D}(x_1, x_2) + \frac{1}{\zeta^2} \sum_{i=3}^{14} x_i^2,
    \label{eq:pel_14d}
\end{equation}
with $\zeta = 0.5 \, (\textrm{eV})^{-1/2}$.
Here we consider the system at a temperature of $T = 500$~K.
%
%
The system is discretized over the domain $x_i \in [-1.5, \, 1.5]$, which forms a hypercube in a 14-dimensional space, using equal lattice spacing $\Delta x_i = 0.1$ for $i = 1, 2, \dots, 14$.
Similar to the 2-dimensional system, the energy minima are located at microscopic states $\textrm{A}$: $(-1.1, \, 0.0, \, 0.0, \dots, 0.0)$ and $\textrm{B}$: $(1.1, \, 0.0, \, 0.0, \dots, 0.0)$, while the first-order saddle points are located at $\textrm{S}_1$: $(0.0, 1.0, 0.0, \dots, 0.0)$ and $\textrm{S}_2$: $(0.0, -1.0, 0.0, \dots, 0.0)$.
The transition pathway of interest corresponds to transitions from microstates with $(x_1=-1.1, \, x_2=0.0)$ to microstates with $(x_1=1.1, \, x_2 = 0.0)$, which solely depends on the first two coordinates of the position vectors.
%
%

\begin{figure*}[htbp!]
    \centering
    \subfloat[]{%
        \includegraphics[width=.34\textwidth]{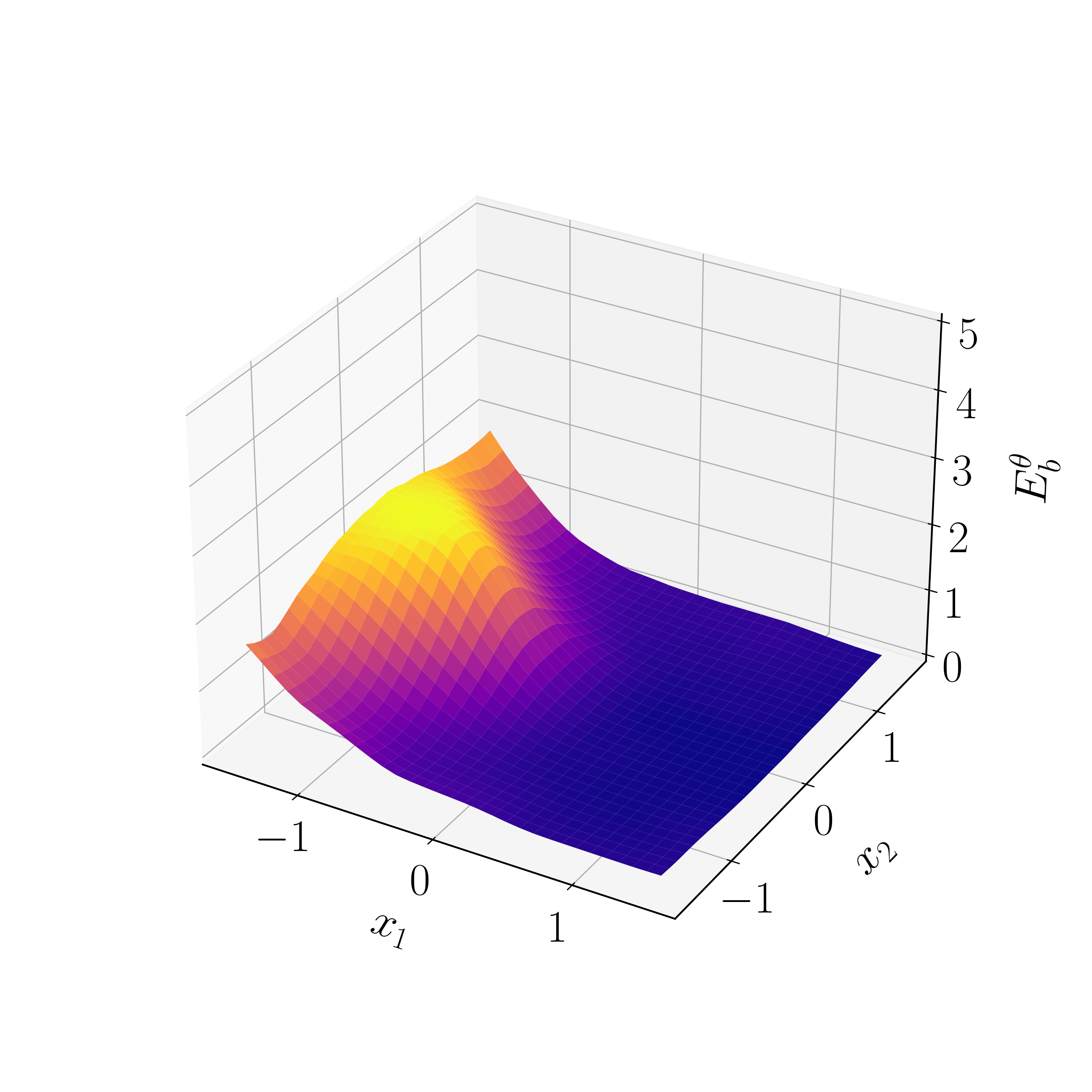}%
        \label{subfig:nngbp_10000e_14d_31n_500K}%
    }
    \subfloat[]{%
        \includegraphics[width=.28\textwidth]{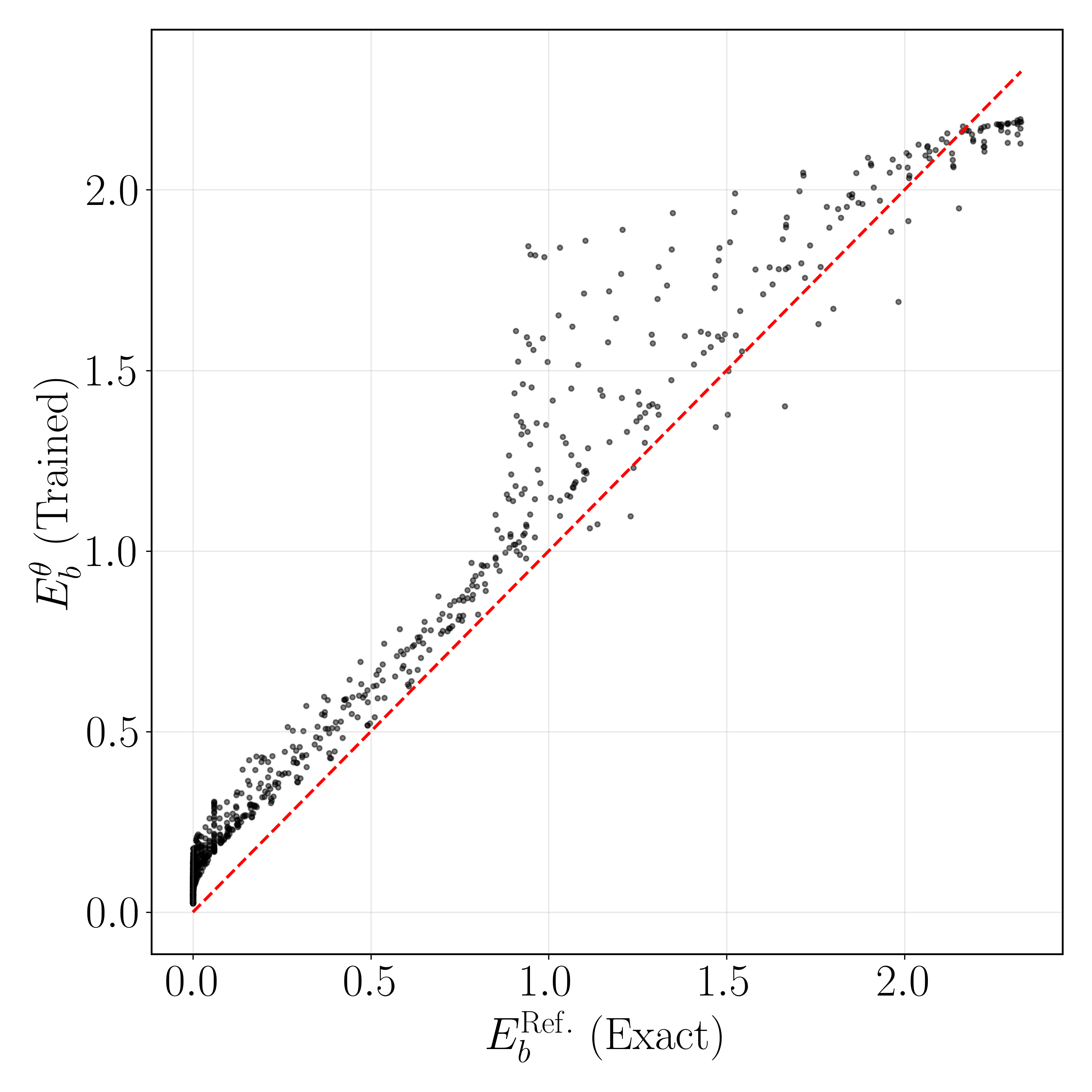}%
        \label{subfig:nngbp_parity_10000e_14d_31n_500K}%
    }
    \subfloat[]{%
        \includegraphics[width=.36\textwidth]{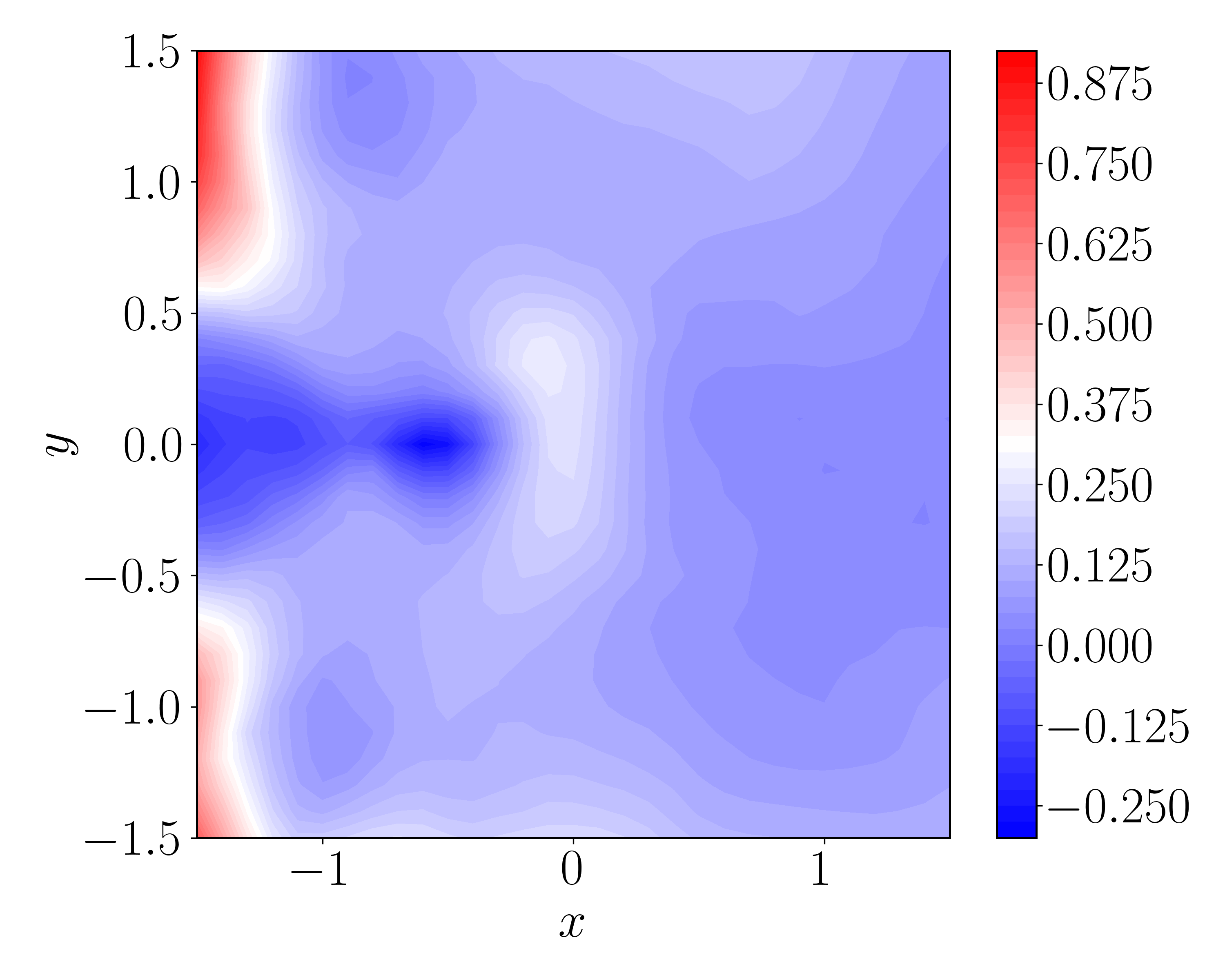}%
        \label{subfig:nngbp_error_10000e_14d_31n_500K}%
    }
    \caption{(a) Neural network bias potential of the 14-dimensional system projected onto a 2-dimensional domain. (b) Parity plot between the neural network bias potential in 14-dimensional system projected onto a 2-dimensional domain and the exact bias potential calculated in 2-dimensional system. (c) Error in neural network bias potential projected onto a 2-dimensional domain compared to the exact bias potential of 2-dimensional system.}
    \label{fig:14d_bias}
\end{figure*}
For the 14-dimensional system, we adopt a hybrid parameterization for the neural network bias potential that can be represented as follows.
\begin{equation}
E_{\textrm{b}}(x_1,x_2,\dots,x_{14};\,\theta) = A \, \exp \left[ -\sum_{i=1}^{14} a_i \, (x_i - c_i)^2 \right] + \textrm{MLP}(x_1,x_2,\dots,x_{14}; \, \theta),
\end{equation}
which is the sum of Gaussian bias potential and a fully connected feed-forward neural network (i.e., $\textrm{MLP}$) and $\{A, a_i, c_i, \theta\}$ form the trainable parameter set. 
The explicit Gaussian term facilitates rapid exploration of the microscopic states during the early stages of the adaptive training scheme.
At the same time, the feed-forward neural network serves as a flexible active correction to capture the complex, non-Gaussian features of the optimal bias potential. 
The MLP comprises two hidden layers, each containing 100 neurons with \texttt{ReLU} activation functions. 
The neural network bias potential is trained directly at the target temperature of $T=500$~K for $10,000$ epochs, using the ADAM~\cite{kingma2014adam} optimizer with a learning rate of $10^{-3}$.
To visualize the neural network bias potential, we project the trained bias potential onto the primary 2-dimensional subspace ($x_1, x_2$) by setting the harmonic coordinates $x_{j \ge 3} = 0$, as shown in Figure~\ref{subfig:nngbp_10000e_14d_31n_500K}. 
Figure~\ref{subfig:nngbp_parity_10000e_14d_31n_500K} shows the comparison between the projected 14-dimensional neural network bias potential and the exact optimal bias potential in the 2-dimensional system.
The spatial distribution of the residual error is plotted in Figure~\ref{subfig:nngbp_error_10000e_14d_31n_500K}, demonstrating that deviations from the exact potential remain minimal within the regions traversed by the dominant transition pathways.

\begin{figure*}[htbp!]
    \centering
    \subfloat[]{%
        \includegraphics[width=.47\textwidth]{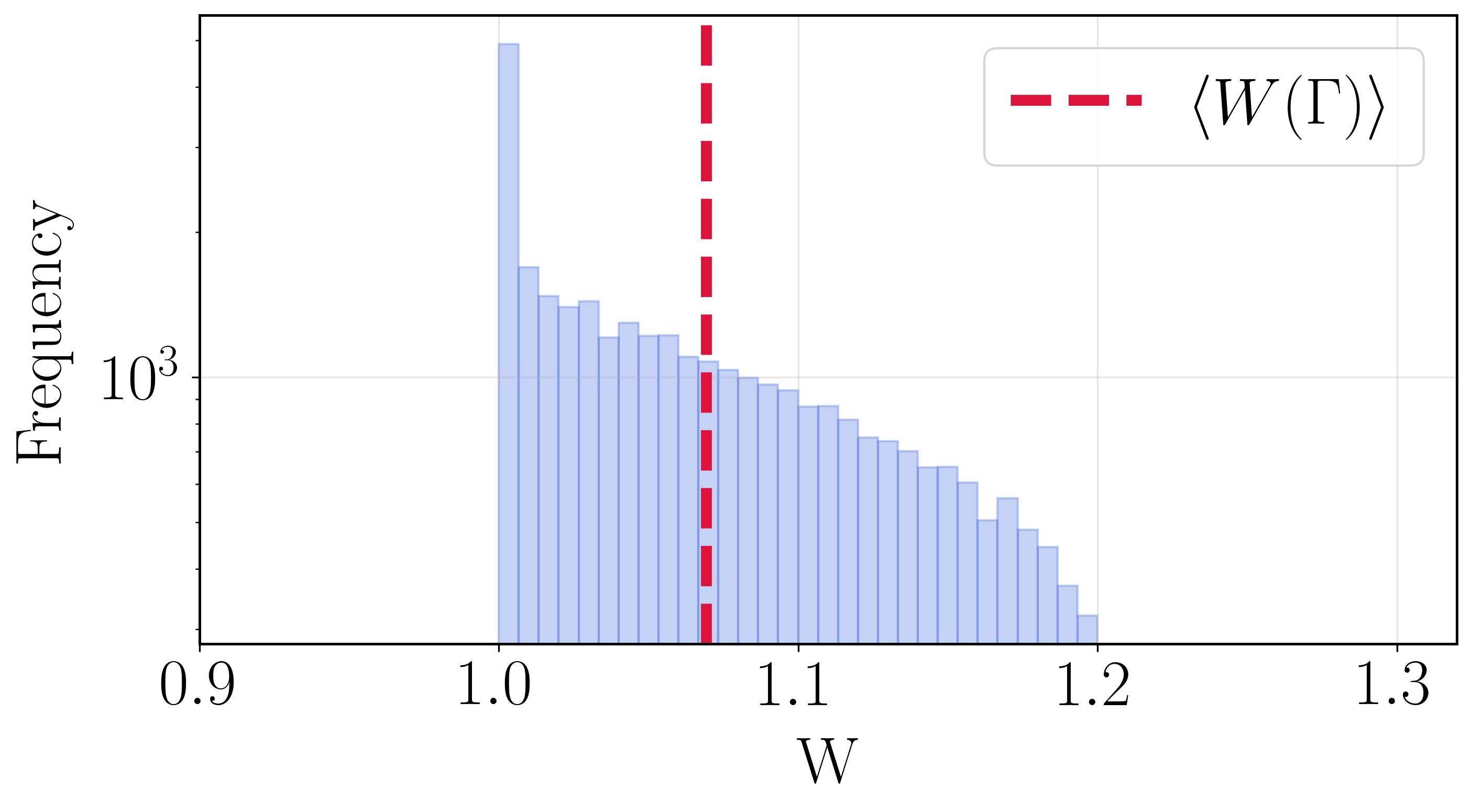}%
        \label{subfig:w_distribution_31n_14d_500.0K_1.0w_1.2w}%
    }
    \hfill
    \subfloat[]{%
        \includegraphics[width=.47\textwidth]{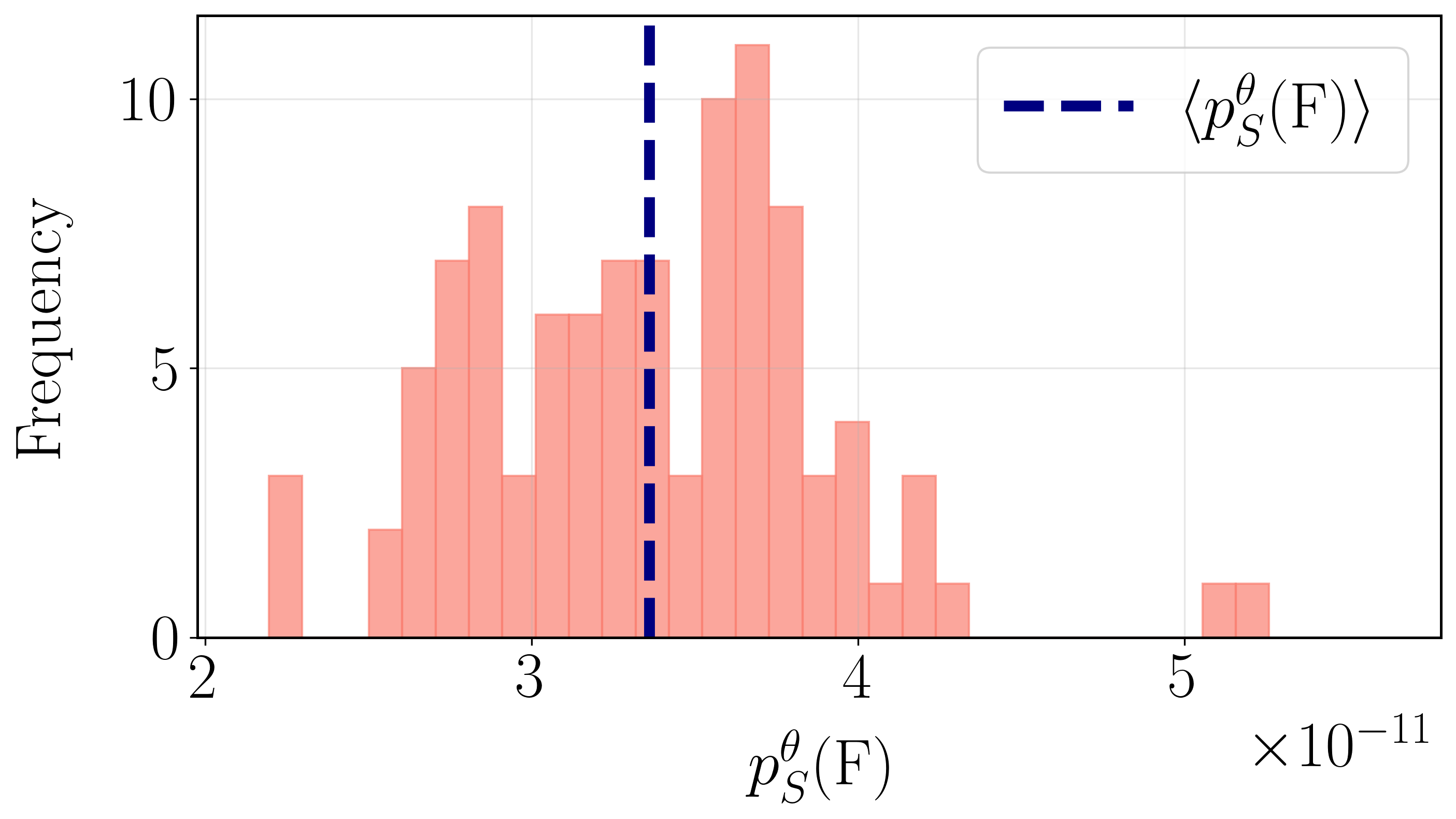}%
        \label{subfig:ps_F_distribution_31n_14d_500K_1.0w_1.2w}%
    }
    \caption{Distribution of (a) weights and (b) success probability estimates by the successful paths sampled using neural network bias potential in 14-dimensional system.}
    \label{fig:14d_w_ps_F}
\end{figure*}
%
%
%
%
Figure~\ref{subfig:w_distribution_31n_14d_500.0K_1.0w_1.2w} presents the weight distribution of successful paths sampled in the 14-dimensional system using the neural network bias potential. 
During the sampling process, the branching random walk (BRW) algorithm is employed to restrict path weights to the interval $[1.0, \, 1.2]$. 
The corresponding distribution of success probability estimates, each derived from an batch of 100 paths, is shown in Figure~\ref{subfig:ps_F_distribution_31n_14d_500K_1.0w_1.2w}. 
Using the unbiased estimator defined in Eq.~(\ref{eq:PS_F_by_imp_sample_R_W}), we obtain a mean success probability $\langle p_{\textrm{S}}(\textrm{F}) \rangle$ of $3.3598 \pm 0.0322 \times 10^{-11}$. 
Concurrently, unbiased Monte Carlo simulations as described in Section~\ref{sec:rate_estimate} yield a mean failure path duration $\langle t_{\textrm{FF}} \rangle$ of $4.9805 \pm 0.0564$. 
Combining these results, the transition rate $r_{\textrm{FS}}$ is estimated as $6.7459 \pm 0.1429 \times 10^{-12}$. 
This value exhibits excellent agreement with the exact transition rate of $6.9191 \times 10^{-12}$ calculated for the underlying 2-dimensional system. 
To highlight the critical importance of the reweighting procedure, we compare this with the estimate obtained by neglecting the path scores following Eq.~(\ref{eq:PS_F_no_imp_sample_R_W}). 
This approximation yields a transition rate of $2.0156 \times 10^{-12}$, which is an underestimation by a factor of three. 
This significant discrepancy demonstrates that 
importance sampling of the transition paths after the approximate bias potential (or committor function) is trained
is crucial for the unbiased estimate of
the transition rate estimate.

%
We further demonstrate that the accelerated sampling framework accurately captures the relative transition probabilities associated with distinct reaction channels, even in high-dimensional spaces. 
The accelerated Monte Carlo simulations estimate the fraction of the total transition flux passing through saddle point S$_1$ to be $0.2850$. 
This value is in close agreement with the prediction of $0.2938$ derived from Kramers' rate theory for the 2-dimensional system. 
Crucially, this agreement reaffirms that the method correctly captures the difference in the entropic pre-factors ($\nu_1$ and $\nu_2$) arising from the local curvature of the potential energy surface at the saddle points (S$_1$ and S$_2$), in addition to the energy barrier difference of the two saddle points.

\section{Conclusion}
\label{sec:conclusion}
We developed a machine learning method for importance sampling of {rare events} in the discrete domain Monte Carlo simulations.
We trained a neural network to learn the optimal bias potential, which is proportional to the log of the optimal importance function. 
Working with the bias potential allows the neural network to deal with extremely small values of the importance function inherent to rare event problems. 
We further demonstrated how this learned bias potential can be used to accurately compute the transition rates between metastable states.
The method is rigorously validated on 2D and 14D model potentials, showing excellent agreement with available theoretical predictions.
%
%
%
This approach provides a unique combination of importance sampling and trainable bias potentials applied to Markov chain Monte Carlo simulations in the discrete domain.
This work demonstrates a novel way to accelerate rare events simulations in high-dimensional and low-temperature systems where conventional methods are intractable.
%
%
%
%
The next logical step in this line of research is applying this framework to atomistic systems, to enable the simulation for long-term processes, such as defect evolution in crystals and protein dynamics in solution.

%
%
%
%
%

\pagebreak

\printbibliography

\pagebreak

\appendix

\section{Kramers' Rate Theory}
\label{appdx:eyring_kramers_theory}

%

Kramers' rate theory predicts the transition rate from the basin containing A through the neighborhood of saddle point S$_1$ to the basin containing B is,
\begin{equation}
    r(\textrm{A} \rightarrow \textrm{S}_1 \rightarrow \textrm{B}) = \mu \, k_{\textrm{B}}T \, \frac{Z(\textrm{S}_1)}{Z(\textrm{A}) \, Z^{\dagger}(\textrm{S}_1)} \, \exp \left[ -\frac{E(\textrm{S}_1) - E(\textrm{A})}{k_{\textrm{B}}T} \right].
    \label{eq:eyring_kramers_theory}
\end{equation}
where $Z(\cdot)$ denotes the local partition functions defined as follows.
%
%
%
%
\begin{equation}
\begin{aligned}
Z(\textrm{A}) &= \int_{\Omega_{\textrm{A}}} \exp \left( -\frac{E(\textbf{x}) - E(\textrm{A})}{k_{\textrm{B}}T} \right) \, \mathrm{d}\textbf{x},\\
Z(\textrm{S}_1) &= \int_{\Omega_{\textrm{S}}^{\perp}} \exp \left( -\frac{E(\textbf{x}) - E(\textrm{S}_1)}{k_{\textrm{B}}T} \right) \, \mathrm{d}\textbf{x},\\
Z^{\dagger}(\textrm{S}_1) &= \int_{\Gamma_{1}} \exp \left( \frac{E(\textbf{x}) - E(\textrm{S}_1)}{k_{\textrm{B}}T} \right) \, \mathrm{d}\textbf{x},
\end{aligned}
\end{equation}
where $\Omega_{\textrm{A}}$ is the basin containing microscopic state A, $\Omega_{\textrm{S}_1}^{\perp}$ denotes the $(d-1)$-dimensional hyperplane orthogonal to the reaction path at the saddle point S$_1$, and $\Gamma_{1}$ represents the 1-dimensional minimum energy path passing through S$_1$.
Comparing Eq.~(\ref{eq:eyring_kramers_theory}) with Eq.~(\ref{eq:Erying_Kramers_r1}), we have,
\begin{equation}
    \nu_1 = \mu \, k_{\textrm{B}}T \, \frac{Z(\textrm{S}_1)}{Z(\textrm{A}) \, Z^{\dagger}(\textrm{S}_1)}.
    \label{eq:eyring_kramers_nu}
\end{equation}
where $\mu$ is related to $\nu_0$ through in Eq.~(\ref{eq:nu_0_mu_kBT_Dx2}).
If one invokes the harmonic approximation, the prefactor $\nu_1$ reduces to the more familiar form,
\begin{equation}
    \nu_1 = \frac{\vert \lambda^*(\textrm{S}_1) \vert}{2\pi} \sqrt{\frac{\textrm{det}(\textbf{H}({\rm A}))}{\vert \textrm{det}(\textbf{H}(\textrm{S}_1))\vert}}.
    \label{eq:eyring_kramers_nu_simp}
\end{equation}
where $\textbf{H}(\cdot)$ is the local Hessian matrix around a given microscopic state and $\lambda^*(\rm S_1)$ is the negative eigenvalue of $\textbf{H}(\rm S_1)$.
In Section~\ref{subsec:transition_rate_mechanisms}, we use the more rigorous expression, Eq.~(\ref{eq:eyring_kramers_nu}), instead of Eq.~(\ref{eq:eyring_kramers_nu_simp}).
We note that the more rigorous expression, Eq.~(\ref{eq:eyring_kramers_nu}), instead of Eq.~(\ref{eq:eyring_kramers_nu_simp}). is used to evaluate the Kramers' rate prediction in Section~\ref{subsec:transition_rate_mechanisms}.

\end{document}